\input harvmac
\input epsf

\def\hat{\widehat}

%
\let\includefigures=\iftrue
%
%
%
\newfam\black
\input rotate
\input epsf
\noblackbox
%
%
\includefigures
\message{If you do not have epsf.tex (to include figures),}
\message{change the option at the top of the tex file.}
\def\figin{\epsfcheck\figin}\def\figins{\epsfcheck\figins}
\def\epsfcheck{\ifx\epsfbox\UnDeFiNeD
\message{(NO epsf.tex, FIGURES WILL BE IGNORED)}
\gdef\figin##1{\vskip2in}\gdef\figins##1{\hskip.5in}
\else\message{(FIGURES WILL BE INCLUDED)}%
\gdef\figin##1{##1}\gdef\figins##1{##1}\fi}
\def\DefWarn#1{}
\def\N{{\cal N}}
\def\figinsert{\goodbreak\midinsert}
\def\ifig#1#2#3{\DefWarn#1\xdef#1{fig.~\the\figno}
\writedef{#1\leftbracket fig.\noexpand~\the\figno}%
\figinsert\figin{\centerline{#3}}\medskip\centerline{\vbox{\baselineskip12pt
\advance\hsize by -1truein\noindent\footnotefont{\bf
Fig.~\the\figno:} #2}}
\bigskip\endinsert\global\advance\figno by1}
\else
\def\ifig#1#2#3{\xdef#1{fig.~\the\figno}
\writedef{#1\leftbracket fig.\noexpand~\the\figno}%
\global\advance\figno by1} \fi
\def\hat{\widehat}
\def\tilde{\widetilde}

\def\yboxit#1#2{\vbox{\hrule height #1 \hbox{\vrule width #1
\vbox{#2}\vrule width #1 }\hrule height #1 }}
\def\fillbox#1{\hbox to #1{\vbox to #1{\vfil}\hfil}}
\def\ybox{{\lower 1.3pt \yboxit{0.4pt}{\fillbox{8pt}}\hskip-0.2pt}}

\def\da{{\dot\alpha}}

\def\rightarrowbox#1#2{
  \setbox1=\hbox{\kern#1{${ #2}$}\kern#1}
  \,\vbox{\offinterlineskip\hbox to\wd1{\hfil\copy1\hfil}
    \kern 3pt\hbox to\wd1{\rightarrowfill}}}

\def\tilp{\tilde\partial}

\def\half{{1\over 2}}
\def\Tr{{{\rm Tr~ }}}

\def\Im{{\rm Im\hskip0.1em}}

\def\vev#1{\langle{#1}\rangle}

\def\tilde{\widetilde}

\def\II{\relax{I\kern-.10em I}}

\def\bar{\overline}

\def\IZ{\relax\ifmmode\mathchoice
{\hbox{\cmss Z\kern-.4em Z}}{\hbox{\cmss Z\kern-.4em Z}}
{\lower.9pt\hbox{\cmsss Z\kern-.4em Z}} {\lower1.2pt\hbox{\cmsss
Z\kern-.4em Z}}\else{\cmss Z\kern-.4em Z}\fi}
\def\IB{\relax{\rm I\kern-.18em B}}
\def\IC{{\relax\hbox{$\inbar\kern-.3em{\rm C}$}}}
\def\ID{\relax{\rm I\kern-.18em D}}
\def\IE{\relax{\rm I\kern-.18em E}}
\def\IF{\relax{\rm I\kern-.18em F}}
\def\IG{\relax\hbox{$\inbar\kern-.3em{\rm G}$}}
\def\IGa{\relax\hbox{${\rm I}\kern-.18em\Gamma$}}
\def\IH{\relax{\rm I\kern-.18em H}}
\def\II{\relax{\rm I\kern-.18em I}}
\def\IK{\relax{\rm I\kern-.18em K}}
\def\IN{\relax{\rm I\kern-.18em N}}
\def\IP{\relax{\rm I\kern-.18em P}}

%
\def\inbar{\,\vrule height1.5ex width.4pt depth0pt}

\font\cmss=cmss10 \font\cmsss=cmss10 at 7pt
\def\IR{\relax{\rm I\kern-.18em R}}

\def\lp10{l_P^{10}}
\def\lp11{l_P^{11}}
\def\R11{R_{11}}

\def\a{\alpha}
\def\b{\beta}
\def\lt{\tilde\lambda}

\def\gb#1{ {\langle #1 ] } }

\def\slashed#1{#1 \!\!\! /}

\def\K{t}

\newbox\tmpbox\setbox\tmpbox\hbox{\abstractfont
}
 \Title{\vbox{\baselineskip12pt\hbox to\wd\tmpbox{\hss
 hep-th/0412103} }}
 {\vbox{\centerline{Generalized Unitarity and One-Loop Amplitudes}
 \bigskip
 \centerline{in $\N=4$ Super-Yang-Mills}
 }}
\smallskip
\centerline{Ruth Britto, Freddy Cachazo and Bo Feng}
\smallskip
\bigskip
\centerline{\it School of Natural Sciences, Institute for Advanced
Study, Princeton NJ 08540 USA}
\bigskip
\vskip 1cm \noindent

\input amssym.tex

One-loop amplitudes of gluons in $\N=4$ gauge theory can be
written as linear combinations of known scalar box integrals with
coefficients that are rational functions. In this paper we show
how to use generalized unitarity to basically read off the
coefficients. The generalized unitarity cuts we use are quadruple
cuts. These can be directly applied to the computation of
four-mass scalar integral coefficients, and we explicitly present
results in next-to-next-to-MHV amplitudes. For scalar box
functions with at least one massless external leg we show that by
doing the computation in signature $(--++)$ the coefficients can
also be obtained from quadruple cuts, which are not useful in Minkowski
signature. As examples, we reproduce the coefficients of some one-,
two-, and three-mass scalar box integrals of the seven-gluon
next-to-MHV amplitude,
and we compute several classes of three-mass and two-mass-hard 
coefficients of next-to-MHV amplitudes to all multiplicities.

\Date{December 2004}

\lref\BernZX{ Z.~Bern, L.~J.~Dixon, D.~C.~Dunbar and
D.~A.~Kosower, ``One Loop N Point Gauge Theory Amplitudes,
Unitarity And Collinear Limits,'' Nucl.\ Phys.\ B {\bf 425}, 217
(1994), hep-ph/9403226.
}

\lref\BernCG{ Z.~Bern, L.~J.~Dixon, D.~C.~Dunbar and
D.~A.~Kosower, ``Fusing Gauge Theory Tree Amplitudes into Loop
Amplitudes,'' Nucl.\ Phys.\ B {\bf 435}, 59 (1995),
hep-ph/9409265.
}

\lref\WittenNN{
E.~Witten,
``Perturbative gauge theory as a string theory in twistor space,''
Commun.\ Math.\ Phys.\  {\bf 252}, 189 (2004)
[arXiv:hep-th/0312171].
}

\lref\CachazoKJ{
F.~Cachazo, P.~Svrcek and E.~Witten,
``MHV vertices and tree amplitudes in gauge theory,''
JHEP {\bf 0409}, 006 (2004)
[arXiv:hep-th/0403047].
}

\lref\BerkovitsJJ{
N.~Berkovits and E.~Witten,
``Conformal supergravity in twistor-string theory,''
JHEP {\bf 0408}, 009 (2004)
[arXiv:hep-th/0406051].
}

\lref\penrose{R. Penrose, ``Twistor Algebra,'' J. Math. Phys. {\bf
8} (1967) 345.}

\lref\berends{F. A. Berends, W. T. Giele and H. Kuijf, ``On
Relations Between Multi-Gluon And Multi-Graviton Scattering,"
Phys. Lett {\bf B211} (1988) 91.}

\lref\berendsgluon{F. A. Berends, W. T. Giele and H. Kuijf,
``Exact and Approximate Expressions for Multigluon Scattering,"
Nucl. Phys. {\bf B333} (1990) 120.}

\lref\bernplusa{Z. Bern, L. Dixon and D. A. Kosower, ``New QCD
Results From String Theory,'' in {\it Strings '93}, ed. M. B.
Halpern et. al. (World-Scientific, 1995), hep-th/9311026.}

\lref\bernplusb{Z. Bern, G. Chalmers, L. J. Dixon and D. A.
Kosower, ``One Loop $N$ Gluon Amplitudes with Maximal Helicity
Violation via Collinear Limits," Phys. Rev. Lett. {\bf 72} (1994)
2134.}

\lref\bernfive{Z. Bern, L. J. Dixon and D. A. Kosower, ``One Loop
Corrections to Five Gluon Amplitudes," Phys. Rev. Lett {\bf 70}
(1993) 2677.}

\lref\bernfourqcd{Z.Bern and  D. A. Kosower, "The Computation of
Loop Amplitudes in Gauge Theories," Nucl. Phys.  {\bf B379,}
(1992) 451.}

\lref\cremmerlag{E. Cremmer and B. Julia, ``The $N=8$ Supergravity
Theory. I. The Lagrangian," Phys. Lett.  {\bf B80} (1980) 48.}

\lref\cremmerso{E. Cremmer and B. Julia, ``The $SO(8)$
Supergravity," Nucl. Phys.  {\bf B159} (1979) 141.}

\lref\dewitt{B. DeWitt, "Quantum Theory of Gravity, III:
Applications of Covariant Theory," Phys. Rev. {\bf 162} (1967)
1239.}

\lref\dunbarn{D. C. Dunbar and P. S. Norridge, "Calculation of
Graviton Scattering Amplitudes Using String Based Methods," Nucl.
Phys. B {\bf 433,} 181 (1995), hep-th/9408014.}

\lref\ellissexton{R. K. Ellis and J. C. Sexton, "QCD Radiative
corrections to parton-parton scattering," Nucl. Phys.  {\bf B269}
(1986) 445.}

\lref\gravityloops{Z. Bern, L. Dixon, M. Perelstein, and J. S.
Rozowsky, ``Multi-Leg One-Loop Gravity Amplitudes from Gauge
Theory,"  hep-th/9811140.}

\lref\kunsztqcd{Z. Kunszt, A. Singer and Z. Tr\'{o}cs\'{a}nyi,
``One-loop Helicity Amplitudes For All $2\rightarrow2$ Processes
in QCD and ${\cal N}=1$ Supersymmetric Yang-Mills Theory,'' Nucl.
Phys.  {\bf B411} (1994) 397, hep-th/9305239.}

\lref\mahlona{G. Mahlon, ``One Loop Multi-photon Helicity
Amplitudes,'' Phys. Rev.  {\bf D49} (1994) 2197, hep-th/9311213.}

\lref\mahlonb{G. Mahlon, ``Multi-gluon Helicity Amplitudes
Involving a Quark Loop,''  Phys. Rev.  {\bf D49} (1994) 4438,
hep-th/9312276.}

\lref\klt{H. Kawai, D. C. Lewellen and S.-H. H. Tye, ``A Relation
Between Tree Amplitudes of Closed and Open Strings," Nucl. Phys.
{B269} (1986) 1.}

\lref\pppmgr{Z. Bern, D. C. Dunbar and T. Shimada, ``String Based
Methods In Perturbative Gravity," Phys. Lett.  {\bf B312} (1993)
277, hep-th/9307001.}

\lref\GiombiIX{ S.~Giombi, R.~Ricci, D.~Robles-Llana and
D.~Trancanelli, ``A Note on Twistor Gravity Amplitudes,''
hep-th/0405086.
}

\lref\WuFB{ J.~B.~Wu and C.~J.~Zhu, ``MHV Vertices and Scattering
Amplitudes in Gauge Theory,'' hep-th/0406085.
}

\lref\Feynman{R.P. Feynman, Acta Phys. Pol. 24 (1963) 697, and in
{\it Magic Without Magic}, ed. J. R. Klauder (Freeman, New York,
1972), p. 355.}

\lref\Peskin{M.~E. Peskin and D.~V. Schroeder, {\it An Introduction
to Quantum Field Theory} (Addison-Wesley Pub. Co., 1995).}

\lref\parke{S. Parke and T. Taylor, ``An Amplitude For $N$ Gluon
Scattering,'' Phys. Rev. Lett. {\bf 56} (1986) 2459; F. A. Berends
and W. T. Giele, ``Recursive Calculations For Processes With $N$
Gluons,'' Nucl. Phys. {\bf B306} (1988) 759. }

\lref\BrandhuberYW{ A.~Brandhuber, B.~Spence and G.~Travaglini,
``One-Loop Gauge Theory Amplitudes In N = 4 Super Yang-Mills From
MHV Vertices,'' hep-th/0407214.
}

\lref\CachazoZB{ F.~Cachazo, P.~Svr\v cek and E.~Witten, ``Twistor
space structure of one-loop amplitudes in gauge theory,''
hep-th/0406177.
}

\lref\passarino{ L.~M. Brown and R.~P. Feynman, ``Radiative Corrections To Compton Scattering,'' Phys. Rev. 85:231
(1952); G.~Passarino and M.~Veltman, ``One Loop Corrections For E+ E- Annihilation Into Mu+ Mu- In The Weinberg
Model,'' Nucl. Phys. B160:151 (1979);
G.~'t Hooft and M.~Veltman, ``Scalar One Loop Integrals,'' Nucl. Phys. B153:365 (1979); R.~G.~
Stuart, ``Algebraic Reduction Of One Loop Feynman Diagrams To Scalar Integrals,'' Comp. Phys. Comm. 48:367 (1988); R.~G.~Stuart and A.~Gongora, ``Algebraic Reduction Of One Loop Feynman Diagrams To Scalar Integrals. 2,'' Comp. Phys. Comm. 56:337 (1990).}

\lref\neerven{ W. van Neerven and J. A. M. Vermaseren, ``Large Loop Integrals,'' Phys. Lett.
137B:241 (1984)}

\lref\melrose{ D.~B.~Melrose, ``Reduction Of Feynman Diagrams,'' Il Nuovo Cimento 40A:181 (1965); G.~J.~van Oldenborgh and J.~A.~M.~Vermaseren, ``New Algorithms For One Loop Integrals,'' Z. Phys. C46:425 (1990);
G.J. van Oldenborgh,  PhD Thesis, University of Amsterdam (1990);
A. Aeppli, PhD thesis, University of Zurich (1992).}

\lref\bernTasi{Z.~Bern, hep-ph/9304249, in {\it Proceedings of
Theoretical Advanced Study Institute in High Energy Physics (TASI
92)}, eds. J. Harvey and J. Polchinski (World Scientific, 1993). }

\lref\morgan{ Z.~Bern and A.~G.~Morgan, ``Supersymmetry relations
between contributions to one loop gauge boson amplitudes,'' Phys.\
Rev.\ D {\bf 49}, 6155 (1994), hep-ph/9312218.
}

\lref\RoiSpV{R.~Roiban, M.~Spradlin and A.~Volovich, ``A Googly
Amplitude From The B-Model In Twistor Space,'' JHEP {\bf 0404},
012 (2004) hep-th/0402016; 
R.~Roiban and A.~Volovich,
``All conjugate-maximal-helicity-violating amplitudes from topological open
string theory in twistor space,''
Phys.\ Rev.\ Lett.\  {\bf 93}, 131602 (2004)
[arXiv:hep-th/0402121],
R.~Roiban, M.~Spradlin and A.~Volovich, ``On The Tree-Level
S-Matrix Of Yang-Mills Theory,'' Phys.\ Rev.\ D {\bf 70}, 026009
(2004) hep-th/0403190,
S.~Gukov, L.~Motl and A.~Neitzke,
``Equivalence of twistor prescriptions for super Yang-Mills,''
arXiv:hep-th/0404085,
I.~Bena, Z.~Bern and D.~A.~Kosower,
``Twistor-space recursive formulation of gauge theory amplitudes,''
arXiv:hep-th/0406133.
}

\lref\CachazoBY{
F.~Cachazo, P.~Svrcek and E.~Witten,
JHEP {\bf 0410}, 077 (2004)
[arXiv:hep-th/0409245].
}

\lref\DixonWI{ L.~J.~Dixon, ``Calculating Scattering Amplitudes
Efficiently,'' hep-ph/9601359.
}

\lref\BernMQ{ Z.~Bern, L.~J.~Dixon and D.~A.~Kosower, ``One Loop
Corrections To Five Gluon Amplitudes,'' Phys.\ Rev.\ Lett.\  {\bf
70}, 2677 (1993), hep-ph/9302280.
}

\lref\berends{F.~A.~Berends, R.~Kleiss, P.~De Causmaecker, R.~Gastmans and T.~T.~Wu, ``Single Bremsstrahlung Processes In Gauge Theories,'' Phys. Lett. {\bf B103} (1981) 124; P.~De
Causmaeker, R.~Gastmans, W.~Troost and T.~T.~Wu, ``Multiple Bremsstrahlung In Gauge Theories At High-Energies. 1. General
Formalism For Quantum Electrodynamics,'' Nucl. Phys. {\bf
B206} (1982) 53; R.~Kleiss and W.~J.~Stirling, ``Spinor Techniques For Calculating P Anti-P $\to$ W+- / Z0 + Jets,'' Nucl. Phys. {\bf
B262} (1985) 235; R.~Gastmans and T.~T. Wu, {\it The Ubiquitous
Photon: Heliclity Method For QED And QCD} Clarendon Press, 1990.}

\lref\xu{Z. Xu, D.-H. Zhang and L. Chang, ``Helicity Amplitudes For Multiple
Bremsstrahlung In Massless Nonabelian Theories,''
 Nucl. Phys. {\bf B291}
(1987) 392.}

\lref\gunion{J.~F. Gunion and Z. Kunszt, ``Improved Analytic Techniques For Tree Graph Calculations And The G G Q
Anti-Q Lepton Anti-Lepton Subprocess,''
Phys. Lett. {\bf 161B}
(1985) 333.}

\lref\GeorgiouBY{ G.~Georgiou, E.~W.~N.~Glover and V.~V.~Khoze,
``Non-MHV Tree Amplitudes In Gauge Theory,'' JHEP {\bf 0407}, 048
(2004), hep-th/0407027.
}

\lref\WuJX{ J.~B.~Wu and C.~J.~Zhu, ``MHV Vertices And Fermionic
Scattering Amplitudes In Gauge Theory With Quarks And Gluinos,''
hep-th/0406146.
}

\lref\WuFB{ J.~B.~Wu and C.~J.~Zhu, ``MHV Vertices And Scattering
Amplitudes In Gauge Theory,'' JHEP {\bf 0407}, 032 (2004),
hep-th/0406085.
}

\lref\GeorgiouWU{ G.~Georgiou and V.~V.~Khoze, ``Tree Amplitudes
In Gauge Theory As Scalar MHV Diagrams,'' JHEP {\bf 0405}, 070
(2004), hep-th/0404072.
}

\lref\Nair{V. Nair, ``A Current Algebra For Some Gauge Theory
Amplitudes," Phys. Lett. {\bf B78} (1978) 464. }

\lref\BernAD{ Z.~Bern, ``String Based Perturbative Methods For
Gauge Theories,'' hep-ph/9304249.
}

\lref\BernKR{ Z.~Bern, L.~J.~Dixon and D.~A.~Kosower,
``Dimensionally Regulated Pentagon Integrals,'' Nucl.\ Phys.\ B
{\bf 412}, 751 (1994), hep-ph/9306240.
}

\lref\CachazoDR{ F.~Cachazo, ``Holomorphic Anomaly Of Unitarity
Cuts And One-Loop Gauge Theory Amplitudes,'' hep-th/0410077.
}

\lref\giel{W. T. Giele and E. W. N. Glover, ``Higher order corrections to jet cross-sections in e+ e- annihilation,'' Phys. Rev. {\bf D46}
(1992) 1980; W. T. Giele, E. W. N. Glover and D. A. Kosower, ``Higher order corrections to jet cross-sections in hadron colliders,'' Nucl.
Phys. {\bf B403} (1993) 633. }

\lref\kuni{Z. Kunszt and D. Soper, ``Calculation of jet cross-sections in hadron collisions at order alpha-s**3,''Phys. Rev. {\bf D46} (1992)
192; Z. Kunszt, A. Signer and Z. Tr\' ocs\' anyi, ``Singular terms of helicity amplitudes at one loop in QCD and the soft limit
of the cross-sections of multiparton processes,'' Nucl. Phys. {\bf
B420} (1994) 550. }

\lref\seventree{F.~A. Berends, W.~T. Giele and H. Kuijf, ``Exact And Approximate Expressions For Multi - Gluon Scattering,'' Nucl. Phys.
{\bf B333} (1990) 120.}

\lref\mangpxu{M. Mangano, S.~J. Parke and Z. Xu, ``Duality And Multi - Gluon Scattering,'' Nucl. Phys. {\bf B298}
(1988) 653.}

\lref\mangparke{M. Mangano and S.~J. Parke, ``Multiparton Amplitudes In Gauge Theories,'' Phys. Rep. {\bf 200}
(1991) 301.}

\lref\grisaru{M. T. Grisaru, H. N. Pendleton and P. van Nieuwenhuizen, ``Supergravity And The S Matrix,'' Phys. Rev.  {\bf D15} (1977) 996; M. T. Grisaru and H. N. Pendleton, ``Some Properties Of Scattering Amplitudes In Supersymmetric Theories,'' Nucl. Phys. {\bf B124} (1977) 81.}

\lref\Bena{I. Bena, Z. Bern, D. A. Kosower and R. Roiban, ``Loops in Twistor Space,'' hep-th/0410054.}

\lref\BernKY{
Z.~Bern, V.~Del Duca, L.~J.~Dixon and D.~A.~Kosower,
``All Non-Maximally-Helicity-Violating One-Loop Seven-Gluon Amplitudes In N =
4 Super-Yang-Mills Theory,''
arXiv:hep-th/0410224.
}

\lref\BrittoNJ{
R.~Britto, F.~Cachazo and B.~Feng,
``Computing one-loop amplitudes from the holomorphic anomaly of unitarity
cuts,''
arXiv:hep-th/0410179.
}

\lref\BidderTX{
S.~J.~Bidder, N.~E.~J.~Bjerrum-Bohr, L.~J.~Dixon and D.~C.~Dunbar,
``N = 1 supersymmetric one-loop amplitudes and the holomorphic anomaly of
unitarity cuts,''
arXiv:hep-th/0410296.
}

\lref\DennerQQ{
A.~Denner, U.~Nierste and R.~Scharf,
``A Compact expression for the scalar one loop four point function,''
Nucl.\ Phys.\ B {\bf 367}, 637 (1991).
}

\lref\BrittoTX{
R.~Britto, F.~Cachazo and B.~Feng,
``Coplanarity in twistor space of N = 4 next-to-MHV one-loop amplitude
coefficients,''
arXiv:hep-th/0411107.
}

\lref\BidderVX{
S.~J.~Bidder, N.~E.~J.~Bjerrum-Bohr, D.~C.~Dunbar and W.~B.~Perkins,
``Twistor Space Structure of the Box Coefficients of N=1 One-loop
Amplitudes,''
arXiv:hep-th/0412023.
}

\lref\BrittoNJ{
R.~Britto, F.~Cachazo and B.~Feng,
``Computing one-loop amplitudes from the holomorphic anomaly of unitarity
cuts,''
arXiv:hep-th/0410179.
}
\lref\KosowerYZ{
D.~A.~Kosower,
``Next-to-maximal helicity violating amplitudes in gauge theory,''
arXiv:hep-th/0406175.
}

\lref\BernKY{
Z.~Bern, V.~Del Duca, L.~J.~Dixon and D.~A.~Kosower,
``All non-maximally-helicity-violating one-loop seven-gluon amplitudes in
N =
4 super-Yang-Mills theory,''
arXiv:hep-th/0410224.
}
                                                                                \lref\sbook{
R. J. Eden, P. V. Landshoff, D. I. Olive and J. C. Polkinghorne,
{\it The Analytic S-Matrix}, Cambridge University Press, 1966.
}

\lref\BernSC{
Z.~Bern, L.~J.~Dixon and D.~A.~Kosower,
``One-loop amplitudes for e+ e- to four partons,''
Nucl.\ Phys.\ B {\bf 513}, 3 (1998)
[arXiv:hep-ph/9708239].
}

\lref\mhvdiag{
C.~J.~Zhu,
``The googly amplitudes in gauge theory,''
JHEP {\bf 0404}, 032 (2004)
[arXiv:hep-th/0403115].
G.~Georgiou and V.~V.~Khoze,
``Tree amplitudes in gauge theory as scalar MHV diagrams,''
JHEP {\bf 0405}, 070 (2004)
[arXiv:hep-th/0404072];
J.~B.~Wu and C.~J.~Zhu,
``MHV vertices and scattering amplitudes in gauge theory,''
JHEP {\bf 0407}, 032 (2004)
[arXiv:hep-th/0406085];
I.~Bena, Z.~Bern and D.~A.~Kosower,
``Twistor-space recursive formulation of gauge theory amplitudes,''
arXiv:hep-th/0406133;
J.~B.~Wu and C.~J.~Zhu,
``MHV vertices and fermionic scattering amplitudes in gauge theory with quarks
and gluinos,''
JHEP {\bf 0409}, 063 (2004)
[arXiv:hep-th/0406146];
G.~Georgiou, E.~W.~N.~Glover and V.~V.~Khoze,
``Non-MHV tree amplitudes in gauge theory,''
JHEP {\bf 0407}, 048 (2004)
[arXiv:hep-th/0407027];
X.~Su and J.~B.~Wu,
``Six-quark amplitudes from fermionic MHV vertices,''
arXiv:hep-th/0409228.
}

\lref\structure{
R.~Roiban, M.~Spradlin and A.~Volovich, ``A Googly
Amplitude From The B-Model In Twistor Space,'' JHEP {\bf 0404},
012 (2004) hep-th/0402016; R.~Roiban and A.~Volovich, ``All Googly
Amplitudes From The $B$-Model In Twistor Space,'' hep-th/0402121;
R.~Roiban, M.~Spradlin and A.~Volovich, ``On The Tree-Level
S-Matrix Of Yang-Mills Theory,'' Phys.\ Rev.\ D {\bf 70}, 026009
(2004) hep-th/0403190;
S.~Gukov, L.~Motl and A.~Neitzke,
``Equivalence of twistor prescriptions for super Yang-Mills,''
arXiv:hep-th/0404085;
S.~Giombi, R.~Ricci, D.~Robles-Llana and D.~Trancanelli,
``A note on twistor gravity amplitudes,''
JHEP {\bf 0407}, 059 (2004)
[arXiv:hep-th/0405086];
I.~Bena, Z.~Bern and D.~A.~Kosower,
``Twistor-space recursive formulation of gauge theory amplitudes,''
arXiv:hep-th/0406133.
F.~Cachazo, P.~Svrcek and E.~Witten,
``Twistor space structure of one-loop amplitudes in gauge theory,''
JHEP {\bf 0410}, 074 (2004)
[arXiv:hep-th/0406177].
}

\lref\looptwistor{
S.~J.~Bidder, N.~E.~J.~Bjerrum-Bohr, L.~J.~Dixon and D.~C.~Dunbar,
``N = 1 supersymmetric one-loop amplitudes and the holomorphic anomaly of
unitarity cuts,''
arXiv:hep-th/0410296;
R.~Britto, F.~Cachazo and B.~Feng,
``Coplanarity in twistor space of N = 4 next-to-MHV one-loop amplitude
coefficients,''
arXiv:hep-th/0411107;
S.~J.~Bidder, N.~E.~J.~Bjerrum-Bohr, D.~C.~Dunbar and W.~B.~Perkins,
``Twistor Space Structure of the Box Coefficients of N=1 One-loop
Amplitudes,''
arXiv:hep-th/0412023.
}

\lref\loopmhv{
A.~Brandhuber, B.~Spence and G.~Travaglini,
``One-loop gauge theory amplitudes in N = 4 super Yang-Mills from MHV
vertices,''
arXiv:hep-th/0407214;
M.~x.~Luo and C.~k.~Wen,
``One-loop maximal helicity violating amplitudes in N = 4 super Yang-Mills
theories,''
JHEP {\bf 0411}, 004 (2004)
[arXiv:hep-th/0410045];
I.~Bena, Z.~Bern, D.~A.~Kosower and R.~Roiban,
``Loops in twistor space,''
arXiv:hep-th/0410054;
M.~x.~Luo and C.~k.~Wen,
``Systematics of one-loop scattering amplitudes in N = 4 super Yang-Mills
theories,''
arXiv:hep-th/0410118;
C.~Quigley and M.~Rozali,
``One-loop MHV amplitudes in supersymmetric gauge theories,''
arXiv:hep-th/0410278;
J.~Bedford, A.~Brandhuber, B.~Spence and G.~Travaglini,
``A twistor approach to one-loop amplitudes in N = 1 supersymmetric Yang-Mills
theory,''
arXiv:hep-th/0410280;
L.~J.~Dixon, E.~W.~N.~Glover and V.~V.~Khoze,
``MHV rules for Higgs plus multi-gluon amplitudes,''
arXiv:hep-th/0411092;
J.~Bedford, A.~Brandhuber, B.~Spence and G.~Travaglini,
``Non-Supersymmetric Loop Amplitudes and MHV Vertices,''
arXiv:hep-th/0412108.
}

\newsec{Introduction}

One-loop amplitudes of gluons in supersymmetric field theories are
four-dimensional cut constructible \refs{\BernZX,\BernCG}. That is
to say, they are completely determined by their unitarity cuts.

In the particular case of $\N=4$ gauge theories, the amplitudes
can be written as a sum over scalar box integrals with rational
functions as coefficients \refs{\BernZX,\BernKR}. Scalar box
integrals are one-loop Feynman integrals in a scalar field theory
with four external legs and four propagators. The scalar in the
loop is massless, and the ones in the external legs could  be
massive. This gives four families classified by the number  of
massive legs. These are called one-, two-, three-, and four-mass
scalar box integrals. Given that all scalar box integrals are
known explicitly \BernKR, the task of computing a given one-loop
amplitude is reduced to that of finding the coefficients.

These coefficients were computed for all maximally helicity
violating amplitudes in \BernZX, for next-to-MHV six-gluon
amplitudes in \BernCG, for next-to-MHV seven-gluon amplitude with
like helicity gluons adjacent in \BrittoNJ\ and for all helicity
configurations in \BernKY.

One common feature of all these computations is that they are
based on the use of unitarity cuts. The basic idea is to compute
the discontinuity of the amplitude across a given branch cut in
the kinematical space of invariants by adding all Feynman graphs
with the same cut. The discontinuity is obtained by ``cutting" two
propagators. This sum of cut Feynman graphs reduces to the product
of two tree-level amplitudes integrated over the Lorentz-invariant phase space
of cut propagators. On the
other hand, each of the scalar box integrals are also Feynman
graphs whose discontinuity can also be computed by cutting two
propagators. Therefore, combining the two different ways of
computing the same discontinuity, one can get information about
the coefficients.

One complication of all these approaches is that there are several
scalar box integrals sharing a given branch cut. Therefore, one
has to disentangle the information of several coefficients at
once. This is done in  \refs{\BernZX,\BernCG,\BernKY} by using
reduction techniques \refs{\passarino,\neerven,\melrose} and in
\refs{\CachazoDR, \BrittoNJ}  by using the holomorphic anomaly
\CachazoBY, which affects the action of differential operators
that test localizations in twistor space \WittenNN.

In this paper we present a different way of computing the
coefficients, using generalized unitarity. Even though several
scalar box integrals can share the same branch cut, it turns out
that their leading singularity is unique.  For a detailed
treatment of generalized unitarity and leading singularities of
Feynman graphs, see Sections 2.9 and 2.2 of \sbook. Therefore, by
studying the discontinuity associated to the leading singularity,
which we denote by $\Delta_{LS}$, we can basically read off a
given coefficient.

For a given Feynman graph, $\Delta_{LS}$ is obtained by cutting
all propagators.  In \BernSC\ the leading singularity of a
three-mass triangle graph, whose discontinuity is computed by 
 a triple cut, was used to
compute its contribution to the $e^+ e^- \rightarrow (\gamma^*,Z)
\rightarrow {\overline q}qgg$ one-loop amplitude. In this paper, we use a quadruple
cut to compute coefficients of four-mass box integrals in one-loop
$\N=4$ gluon amplitudes. We then turn our attention to scalar box
integrals with at least one massless leg. In Minkowski space,
$\Delta_{LS}$ for these box integrals, which is still a quadruple
cut, does not give information about the coefficients, as we
explain in section 2. One has to use at most triple
cuts.\foot{Triple cuts were used in \BernKY\ to conclude that some
coefficients in one-loop $\N=4$ gluon amplitudes must vanish for
special choices of helicity.} One disadvantage of triple cuts is
that there might be several box integrals with the same
singularity, and therefore several unknown coefficients will show
up at once.

We find a way out by going to signature $(--++)$.\foot{Alternatively, it is perhaps more natural to drop the restriction of real momenta and work with the complexified Lorentz group.  We thank E. Witten for pointing this out.}
 We show that in
this signature all scalar box integral coefficients can be
computed by a quadruple cut.
 This allows us to also read off the
coefficients directly,
since again only one coefficient contributes in a given cut. The
quadruple cut integral is completely localized by the four delta
functions of the cut propagators. It reduces to a product of four
tree-level amplitudes, which can be easily computed by MHV
diagrams \refs{\CachazoKJ,\mhvdiag,\KosowerYZ}.  This is a very simple way of computing any
given coefficient in any one-loop $\N=4$ gluon amplitude.

We illustrate this procedure by computing several coefficients of
one-, two-, three- and four-mass box scalar integrals. For the
first three cases we give examples for 
next-to-MHV amplitudes, including some results to all multiplicities, 
and for the last we give examples for
an eight-gluon next-to-next-to-MHV (NNMHV) amplitude.

Motivated by the introduction of twistor string theory \WittenNN,
there has recently been interest in
 the twistor-space localization of gauge theory amplitudes \refs{\CachazoKJ,\structure,\CachazoBY}.
In particular, for supersymmetric gauge theories, 
the coefficients
of box, triangle and bubble scalar integrals also exhibit simple
twistor space structure \refs{\BernKY,\looptwistor}.
At one-loop, this structure can be understood from the MHV diagram formulation 
\loopmhv.

Since no four-mass coefficient has previously been presented in the
literature, we consider its twistor space support.  This is done by studying various differential operators acting on unitarity cuts.

This paper is organized as follows: In section 2, we discuss the
generalized unitarity method in general and discuss how one can
use quadruple cuts for all scalar box integrals in signature
$(--++)$. In section 3, we present examples of each type of scalar box integral, starting with the
four-mass, and then the three-, two-, and one-mass. 
In section 4, we give several examples of infinite classes of coefficients in next-to-MHV amplitudes.
In the
appendices, we consider in detail the discontinuity of the
four-mass scalar box integral associated to branch cuts in all
possible channels and use this to get consistency checks on the
new coefficient obtained in section 3.

\subsec{Preliminaries}

One-loop $\N=4$ amplitudes of gluons can be written as a linear
combination of scalar box integrals with rational coefficients
\refs{\BernKR, \BernZX}. In this paper, we concentrate on the leading-color
contribution, which is the part of the full
amplitude proportional to $N \Tr (T^{a_1}\ldots T^{a_n})$.

We write this schematically as follows:
\eqn\ampl{A_{n:1} = \sum \left( {\hat b} I^{1m} + {\hat c}
I^{2m~e}+ {\hat d}I^{2m~h} + {\hat g} I^{3m}+ {\hat
f}I^{4m}\right).}
The integrals are defined in dimensional regularization as
\eqn\bbo{I_4 = -i(4\pi)^{2-\epsilon} \int
{d^{4-2\epsilon}\ell\over (2\pi)^{4-2\epsilon}} {1\over \ell^2
(\ell-K_1)^2(\ell-K_1-K_2)^2(\ell+K_4)^2}. }
The external momenta $K_i$ are taken to be outgoing and are given by
the sum of the momenta of consecutive external gluons, as shown in
Figure 1. The labels $1m,2m,3m,4m$ refer to the number of legs
$K_n$ such that $K_n^2 \neq 0$, or equivalently, the number of
vertices in the box with more than one external gluon.
 For two-mass box integrals, there are two
inequivalent arrangements of massive legs. Either they are
adjacent ($I^{2m~h}$) or they are diagonally opposite ($I^{2m~e}$). All these integrals are UV finite but suffer from IR
divergences, except for $I^{4m}$ which is finite.

\ifig\kkmbp{Scalar box integrals.  (a) The outgoing external
momenta at each of the vertices are $K_1,K_2,K_3,K_4$, defined to
correspond to sums of the momenta of gluons in the exact
orientation shown. (b) One-mass $I^{1m}_{n:i}$. (c) Two-mass
``easy" $I^{2m~e}_{n:r;i}$. (d) Two-mass ``hard"
$I^{2m~h}_{n:r;i}$. (e) Three-mass $I^{3m}_{n:r:r';i}$. (f)
Four-mass $I^{4m}_{n:r:r':r";i}$. }
{\epsfxsize=0.70\hsize\epsfbox{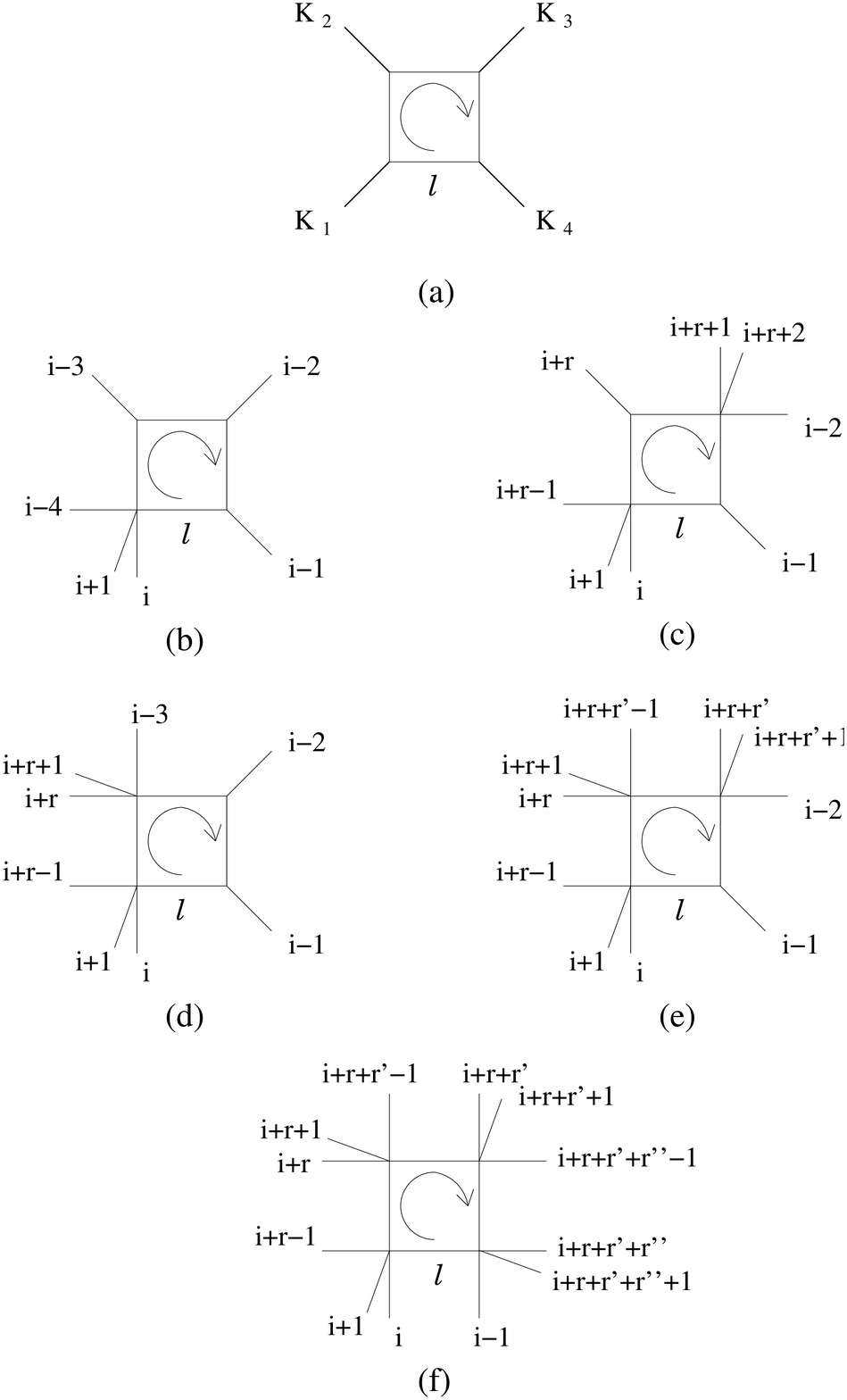}}

From here on, we will drop the dimensional regularization
parameter $\epsilon$, because we will only deal with cuts that are
finite. Moreover, we work in four-dimensional
Minkowski space.

In the literature, it is common to write the amplitude in terms of
scalar box functions. These functions are given in terms of the
scalar box integrals as follows.
\eqn\boxf{ \eqalign{ & I^{1m}_{4:i} = -2{F^{1m}_{n:i}\over K_{41}^2
K_{12}^2}, \quad I^{2m~e}_{4:r;i} = -2{F^{2m~e}_{n:r;i} \over
K_{41}^2 K_{12}^2- K_1^2 K_3^2}, \quad I^{2m~h}_{4:r;i} =
-2{F^{2m~h}_{n:r;i}\over K_{12}^2 K_{41}^2 }, \cr &
I^{3m}_{4:r:r';i} = -2{F^{3m}_{n:r:r';i}\over K_{41}^2 K_{12}^2-
K_1^2 K_3^2}, \quad I^{4m}_{4:r:r':r";i} =
-2{F^{4m}_{n:r:r':r";i}\over K_{41}^2 K_{12}^2\rho}.} }
Here, and throughout the paper, we define
\eqn\defrho{\eqalign{K_{mn} &= K_m+K_n, \cr \rho &=
\sqrt{1-2\lambda_1-2\lambda_2+(\lambda_1-\lambda_2)^2}, \cr
\lambda_1 & = {K_1^2 K_3^2 \over K_{41}^2 K_{12}^2}, \cr \lambda_2
& = {K_2^2 K_4^2 \over K_{41}^2 K_{12}^2}.}}

Then we can alternatively write \ampl\ as a linear combination of
scalar box functions
\eqn\kimo{ A_{n:1} = \sum \left( b F^{1m} + c F^{2m~e}+ d F^{2m~h}
+ g F^{3m}+ f F^{4m}\right) . }
Each way of writing the amplitude has its own advantages and
disadvantages. In \ampl, all coefficients are rational, but
their twistor space support is not simple. In \kimo, the
coefficient of the four-mass box function is not rational, for it
contains a square root, but all coefficients have simple twistor
space structure. For a discussion of the localization in
twistor space of the four-mass scalar box function coefficient, see the appendices.  For the rest of the body of the paper, we will work mainly with scalar box integrals and their coefficients as formulated in \ampl.


\newsec{Generalized Unitarity and Quadruple Cuts}

One-loop amplitudes in field theory have several singularities as
complex functions of the kinematical invariants. In $\N=4$ gauge
theory, the singularities can only be those of the scalar box
integrals \bbo\ and of the coefficients in \ampl. Since the
coefficients are rational functions, they are not affected by
branch cut singularities. Therefore one can get information about
them by studying the discontinuities of the amplitude across the
cuts.

In fact, most of the techniques for computing the coefficients
efficiently are based on studying unitarity cuts 
\refs{\BernZX,\BernCG,\CachazoDR,\BrittoNJ,\BernKY}. The basic
idea is to consider the branch cut singularity of the amplitude in
a given channel. The discontinuity across this branch cut can be
computed by a cut integral on both sides of the equation
\eqn\ampl{A_{n:1} = \sum \left({\hat b} I^{1m} + {\hat c}
I^{2m~e}+ {\hat d}I^{2m~h} + {\hat g} I^{3m}+ {\hat
f}I^{4m}\right).}
On the left hand side, one cuts two propagators
of all Feynman integrals participating in this channel, while on
the right hand side one cuts two propagators of scalar box
functions.

By ``cutting propagators'' we mean the following. In Minkowski
space, propagators in one loop integrals are defined by using
Feynman's $i\epsilon$ prescription, i.e., $1/(P^2+i\epsilon)$.
This is equal to the principal value of $1/P^2$ plus
$i\delta^{(+)} (P^2)$, where $(+)$ indicates a restriction to the
future light-cone. Cutting a propagator means removing the
principal part, i.e., replacing the propagator by $i\delta^{(+)}
(P^2)$.

The sum over cut Feynman integrals on the left of \ampl\ becomes
an integral over a Lorentz invariant phase space of the product of
two tree-level amplitudes. To be more explicit, consider the cut
in the $(i,i+1,...,j)$-channel,
\eqn\cutii{C = \int d\mu~A^{\rm tree} (\ell_1,i,...,j,
\ell_2)~A^{\rm tree}(-\ell_2,j+1,..., i-1,-\ell_1),}
where $d\mu$ is the Lorentz invariant phase space measure for
$(\ell_1,\ell_2)$. It is given explicitly by
\eqn\mme{ d\mu = \delta^{(+)} (\ell_1^2) ~\delta^{(+)} (\ell_2^2)
~\delta^{(4)}(\ell_1 + \ell_2 - P_{ij}),}
where $P_{ij}$ is the sum of the momenta of gluons from $i$ to
$j$.

If we denote the discontinuity across the branch cut of a given
scalar box integral $I$ by $\Delta I$, then we have the following
equation.
\eqn\compa{\eqalign{ & \int d\mu A^{\rm tree} (\ell_1,i,...,j,
\ell_2)A^{\rm tree}(-\ell_2,j+1,..., i-1,-\ell_1) = \cr &
\sum\left( {\hat b} \Delta I^{1m} + {\hat c} \Delta I^{2m~e}+
{\hat d}\Delta I^{2m~h} + {\hat g} \Delta I^{3m}+ {\hat f}\Delta
I^{4m}\right). }}
As discussed in the introduction, these equations have been used
to get the coefficients for MHV amplitudes and six- and seven-
gluon next-to-MHV amplitudes \refs{\BernZX,\BernCG,\CachazoDR,\BrittoNJ,\BernKY}.

However, extracting the coefficients is not a simple task in
general. The main reason is that several scalar box integrals
share a given cut, and therefore their unknown coefficients enter
in the equation at the same time.

Several approaches exist to extract the coefficients. One is based
on reduction techniques that allow writing the integrand of
\cutii\ as a sum of  terms that have the structure of cut scalar
box functions \refs{\BernZX,\BernCG}. Another method
\refs{\CachazoDR,\BrittoNJ}  uses operators that test localization
in twistor space to get rational functions on both sides of the
discontinuity of \ampl\ and compare the pole structure.

It is the aim of this section to use higher order singularities 
to reduce the number of scalar box functions that enter in the
generalization of \compa. Ideally, we would like to find only one
scalar box integral on the right hand side of \compa.

This can easily be done for the four-mass scalar box integral.
Consider the discontinuity associated to its leading singularity,
$\Delta_{LS}$. As mentioned in the introduction, this is computed
from the integral
\eqn\ifou{I^{4m}=\int d^4\ell { 1\over (\ell^2+i\epsilon)
((\ell-K_1)^2 + i\epsilon )((\ell -K_1-K_2)^2 + i\epsilon
)((\ell+K_4)^2 + i\epsilon )} }
by cutting all four propagators:
\eqn\wpuvv{\Delta_{LS} I^{4m} =\int d^4\ell ~\delta^{(+)}(\ell^2)
~\delta^{(+)}((\ell-K_1)^2) ~\delta^{(+)}((\ell
-K_1-K_2)^2)~\delta^{(+)}((\ell+K_4)^2).}

\ifig\druple{A quadruple cut diagram. Momenta in the cut
propagators flows clockwise and external momenta are taken
outgoing.  The tree-level amplitude $A^{\rm tree}_1$, for example, has external momenta $i+1,...,j,\ell_2,\ell_1$.} {\epsfxsize=0.40\hsize\epsfbox{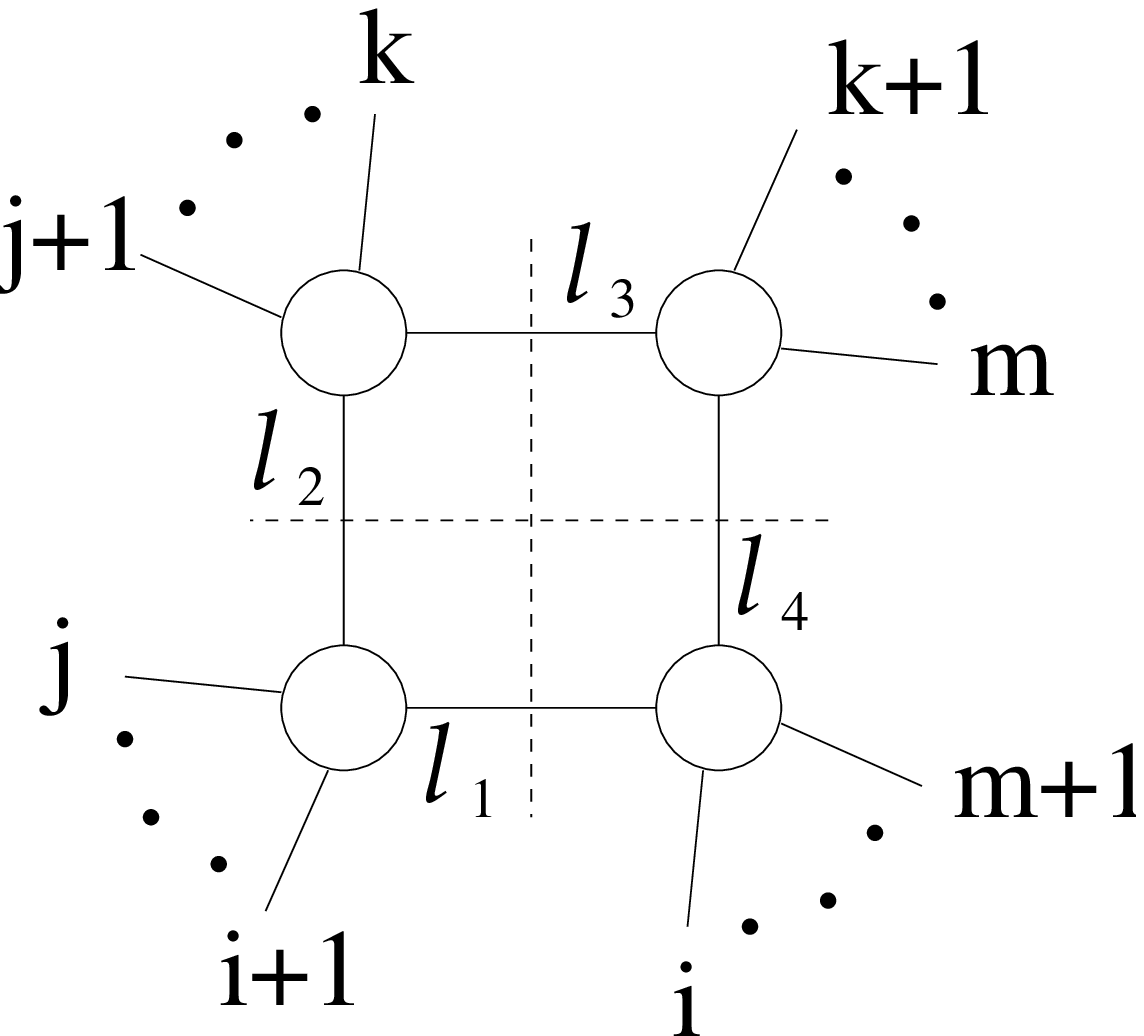}}

Now it turns out that no other box integral in \ampl\ shares the
same singularity. Therefore, the generalization of \compa\ is
\eqn\genfo{\eqalign{ & \int d^4\ell
~\delta^{(+)}(\ell^2)~\delta^{(+)}((\ell-K_1)^2)
~\delta^{(+)}((\ell -K_1-K_2)^2)~\delta^{(+)}((\ell+K_4)^2) \times
\cr &
 A^{\rm tree}_1 A^{\rm tree}_2
 A^{\rm tree}_3 A^{\rm tree}_4 = {\hat f} \Delta_{LS} I^{4m},
}}
where $A^{\rm tree}_n$ is the tree-level amplitude at the corner
with total external momentum $K_n$. See \druple.

Note that since there are four delta functions, and $\ell$ is a
vector in four dimensions, then the integral is localized and
equals a jacobian ${\cal J}=(4 K_{41}^2 K_{12}^2 \rho)^{-1}$. 
Moreover, the same jacobian appears
on both sides of \genfo, and we find that
\eqn\foco{ {\hat f} = {1 \over |{\cal S}|}
\sum_{{\cal S},J} n_J (A^{\rm tree}_1 A^{\rm tree}_2 A^{\rm
tree}_3 A^{\rm tree}_4), }
where the sum is over the possible spins $J$ of  internal particles and the solution set ${\cal S}$ of
 $\ell$'s for
the localization equations, $|{\cal S}|$ is the number of these solutions,
 and $n_J$ is the number of particles of spin $J$.

This equation gives the coefficient $\hat f$ of any four-mass box
function in any one-loop amplitude, as we discuss in more detail
in section 2.1.

\ifig\boxdef{Scalar box integral with at least one massless leg.  We use thick lines for massive legs and thin lines for massless legs.}
{\epsfxsize=0.40\hsize\epsfbox{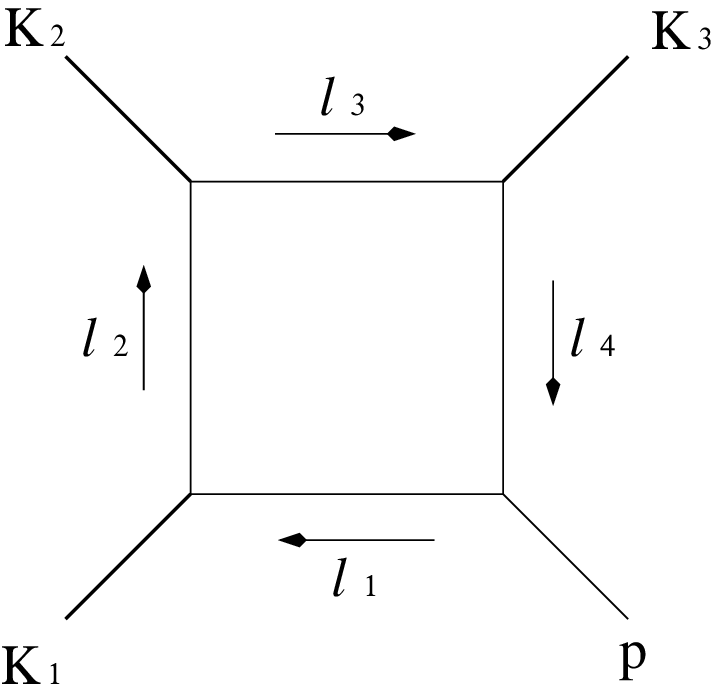}}

The natural question at this point is whether the same thing can
be done for other scalar box integral coefficients. It turns out
that in Minkowski space this is not possible. Recall that all
other scalar box integrals contain at least one massless vertex.
This means that at least one of the four tree-level amplitudes in
the quadruple cut must be a three-gluon amplitude. One problem is
that all three-gluon amplitudes vanish on-shell, as will become
clear in the discussion below.

For definiteness, consider the three-mass scalar box integral in
Figure 3. Let us denote the momenta of the three-gluon vertex by $\ell_1$,
$\ell_4$ and $p$. We have to impose that all three vectors be
lightlike and that $\ell_4 = \ell_1+ p $. From this it is easy to
see that necessary conditions for all three vectors to be
lightlike are $\ell_1\cdot p = \ell_4 \cdot p = \ell_1\cdot
\ell_4 = 0$. In Minkowski signature one can go to a frame where
$p=(E,0,0,E)$. Then it is easy to see from $\ell_1\cdot p =0$ that
$\ell_1 = \alpha p$. Likewise, $\ell_4 = \beta p$. This means that
all three gluons are collinear. Momentum conservation then implies
that $\beta = \alpha +1$.

On the other hand, from $\ell_2 = \ell_1-K_1$ and $\ell_2^2 =0$ we
can compute $\alpha$, and from $\ell_3 = \ell_4 + K_3$ we get
$\beta$. They are given by
\eqn\albe{ \alpha = {K_1^2\over 2K_1\cdot p}, \qquad \beta = -
{K_3^2\over 2K_3\cdot p}. }
Now we can ask if the equation $\beta = \alpha +1$ is satisfied.
It is not difficult to see that this equation is equivalent to
$K_1^2 K_3^2 - (K_1+K_2)^2(K_1+p)^2 = 0$. Note that the left hand
side is precisely the denominator in \boxf. This means that this
singularity is a pole, as opposed to the usual branch cut
singularity. The discontinuity associated to it is a delta
function,\foot{
In one real variable, the statement that the discontinuity associated to a pole is a delta function is expressed in the familiar relation
${1 \over x-i\epsilon}-{1 \over x+i\epsilon}=2\pi i \delta(x)$.
} reflecting the fact that only three of the four delta
functions suffice to perform the integral.

This implies that the quadruple cut equation \genfo\ is in this
case given by
\eqn\trfo{ \int d\mu~ A^{\rm tree}_1 A^{\rm tree}_2 A^{\rm tree}_3
A^{\rm tree}_4 = {\hat g}~ \delta (K_1^2 K_3^2 -
(K_1+K_2)^2(K_1+p)^2).}
Note that the left hand side is trivially zero, for $A_4$ is a
three-gluon tree-level amplitude on-shell which vanishes in
Minkowski space. Then we find that
\eqn\qoqo{{\hat g}~ \delta (K_1^2 K_3^2 - (K_1+K_2)^2(K_1+p)^2) = 0,
}
and therefore the coefficient must have a zero at the support of
the delta function. Indeed, recall that the relation between the
box integral coefficient ${\hat g}$ and the box function
coefficient $g$ is given by ${\hat g} = \half (K_1^2 K_3^2 -
(K_1+K_2)^2(K_1+p)^2) g$.

This proves that the information about $g$ disappears from the
equation, and the quadruple cut cannot be used to learn anything
about $\hat g$, except for the presence of the factor $\half (K_1^2
K_3^2 - (K_1+K_2)^2(K_1+p)^2)$.

If we insist on using Minkowski signature, we have to study triple
cuts.\foot{To be precise, all unitarity cut analyses require complexified momenta.  We thank L. Dixon for this clarification.}  The disadvantage is that in most cases, more than one box
integral will share the same singularity. The situation is better
than with the usual cut in \compa, but it is still complicated in
most cases.

\subsec{Quadruple Cut in Signature $(--++)$}

The main source of the problem for scalar box integrals with at
least one massless leg is the presence of the three-gluon vertex.
The fact that this always vanishes in Minkowski space made
impossible the extraction of their coefficients from the quadruple
cut.

As explained carefully in section 3 of \WittenNN, three-gluon
amplitudes with helicities $(++-)$ and $(--+)$ do not necessarily
vanish in other signatures. In particular, motivated by the study
of a string theory with target twistor space \WittenNN, Witten
considered the three-gluon amplitude in signature $(--++)$. It
turns out that this is precisely what is needed in order to use
the quadruple cuts to read off all coefficients.

We now review the analysis done in \WittenNN. In four dimensions we
can write a null vector as a bispinor, $p_{a\dot a} =
\lambda_a\tilde\lambda_{\dot a}$. The inner product of two null
vectors $p$ and $q$ is given by $2p\cdot q =
\vev{\lambda_p~\lambda_q}[\tilde\lambda_p ~ \tilde\lambda_q]$,
where the brackets represent the natural inner products of spinors
of positive and negative chirality.

A tree-level three-gluon amplitude with helicity $(++-)$ or
$(--+)$ is given respectively by \parke\
\eqn\thee{ A^{\rm tree}_3 (p^+,q^+,r^-) = {[\lt_p~\lt_q]^3\over
[\lt_r~\lt_p][\lt_q~\lt_r]}, \qquad A^{\rm tree}_3 (p^-,q^-,r^+) =
{\vev{\lambda_p~\lambda_q}^3\over \vev{\lambda_r~\lambda_p}\vev{\lambda_q~\lambda_r}}.}
In Minkowski space and for real momenta, $\lambda_p$ and
$\tilde\lambda_p$ are complex but not independent,
$\tilde\lambda_p = \pm {\bar\lambda}_p$. Therefore, when $p\cdot
q= 0 $, this means that both $\vev{p~q}$ and $[p~q]$ vanish. This
implies that both amplitudes in \thee\ also vanish.

On the other hand, in signature $(--++)$ and for real momenta,
both $\lambda_p$ and $\tilde\lambda_p$ are real and independent.
Therefore, $2p\cdot q =\vev{\lambda_p~\lambda_q}[\tilde\lambda_p ~
\tilde\lambda_q]=0$ has two solutions. Either
$\vev{\lambda_p~\lambda_q}=0$ or $[\tilde\lambda_p ~
\tilde\lambda_q]=0$.

Since momentum conservation implies that $p\cdot q = p\cdot r =
q\cdot r = 0$, we find that if $\vev{\lambda_p~\lambda_q}=0$, then
$\vev{\lambda_p~\lambda_r}=0$ and $\vev{\lambda_r~\lambda_q}=0$
must also hold. Therefore all three $\lambda$'s are proportional.
Likewise if we choose $[\tilde\lambda_p~\tilde\lambda_q] =0$, then
all three $\tilde\lambda$'s are proportional.

Now it is clear that if we are faced with a $(++-)$ tree amplitude,
we should choose all $\lambda$'s to be proportional, and then the corresponding
amplitude in \thee\ will not vanish. Likewise, if we are faced
with a $(--+)$, we should choose all $\tilde\lambda$'s
to be proportional.

Having solved the problem of the vanishing of the amplitudes, we
have to deal with the meaning of a ``unitarity cut" in signature
$(--++)$. We certainly cannot offer a formal theory here, but some
interesting developments have appeared in the literature about
field theories in $(--++)$ signature.  For a recent review, 
see \ref\BarsXV{I.~Bars,
``2T physics 2001,''
arXiv:hep-th/0106021.}.

Here we take a more operational approach in order to compute the
coefficients. Since the coefficients are written in terms of
invariant products of spinors, once they are computed they can be
used in Minkowski space.

Consider the quadruple cut measure in Minkowski space. Recall that
each delta function is actually given by
\eqn\twot{ \delta^{(+)}(P^2) = \vartheta(E_P) \delta (P^2), }
where $E_P$ is the zeroth component of $P^\mu$. Both
sides of \genfo\ contain the same measure and in particular the
same factors of $\vartheta(x)$. Since the integral is localized,
one can drop the $\vartheta(x)$ factors on both sides. We then
take as our definition of a quadruple cut in signature $(--++)$
the following:
\eqn\hawy{\eqalign{ & \int d^4\ell~
\delta(\ell^2)~\delta((\ell-K_1)^2) ~\delta((\ell
-K_1-K_2)^2)~\delta((\ell+K_4)^2)
 A^{\rm tree}_1 A^{\rm tree}_2 A^{\rm tree}_3 A^{\rm tree}_4  = \cr & {\hat g}\int d^4\ell~ \delta(\ell^2)~\delta((\ell-K_1)^2)
~\delta((\ell -K_1-K_2)^2)~\delta((\ell+K_4)^2). }}
Here we use $\hat g$ to represent a generic coefficient in \ampl\ with at least one massless leg.

In \hawy\ the delta functions are ordinary Dirac delta functions.
One can solve for $\ell_{a\dot a}$, taking into account that
solutions for which one or more of the tree-level amplitudes
vanish do not contribute. Therefore we find that the coefficient
must be equal to
\eqn\newco{ {\hat g} = {1 \over |{\cal S}|}\sum_{{\cal S},J} n_J A^{\rm tree}_1 A^{\rm tree}_2 A^{\rm
tree}_3 A^{\rm tree}_4,}
where again the sum is over all solutions.

In the rest of the paper, we illustrate this procedure in detail.

\newsec{Examples}

The previous section demonstrated that coefficients of box integrals
may be computed from the formula
\eqn\besatway{{\hat a}_\alpha = {1 \over |{\cal S}|}\sum_{{\cal S},J} n_J A^{\rm tree}_1 A^{\rm tree}_2 A^{\rm tree}_3 A^{\rm tree}_4,
}
where  $J$ is the spin of a particle in the $\N=4$ multiplet and ${\cal S}$ is the set of all  solutions of the on-shell conditions for the internal lines,
\eqn\nullshe{
{\cal S}=\{ ~\ell ~|~ \ell^2=0, \quad (\ell-K_1)^2=0, \quad (\ell
-K_1-K_2)^2=0, \quad (\ell+K_4)^2=0\}.
}

In this section, we present a variety of applications of the formula \besatway.

The explicit covariant solution of \nullshe\ for the vector $\ell$ is given by:
\eqn\ibciz{\eqalign{
\ell &= \beta_1 K_1+ \beta_2 K_2+\beta_3 K_4+\beta_4 P; \cr
P_\mu &= \epsilon_{\mu\nu\rho\sigma} K_1^\nu K_2^\rho K_4^\sigma, \cr
\beta_1 & =  {1\over 2P^2} \left[(K_2\cdot K_4) (-(K_1\cdot K_4)
 (2 (K_1\cdot K_2) + K_2^2) + K_1^2 (K_2\cdot K_4)) \right. \cr &  \left.
 + (2 (K_1\cdot K_2)^2 - (K_1^2 + (K_1\cdot K_4)) K_2^2 +
          (K_1\cdot K_2) (K_2^2 + (K_2\cdot K_4))) K_4^2 \right], \cr
\beta_2 & =   {1\over 2P^2} \left[(K_1\cdot K_4) (2 (K_1\cdot K_2)
(K_1\cdot K_4) + (K_1\cdot K_4) K_2^2 - K_1^2 (K_2\cdot K_4))
\right. \cr &  \left. -(-(K_1\cdot K_2) (K_1\cdot K_4)
+ K_1^2 ((K_1\cdot K_2) + K_2^2 + (K_2\cdot K_4))) K_4^2\right], \cr
\beta_3 & =   {1\over 2P^2} \left[(K_1\cdot K_2) (-(K_1\cdot K_4) K_2^2
+ K_1^2 (K_2\cdot K_4))\right. \cr & \left.
-(K_1\cdot K_2)^2 (2 (K_1\cdot K_4) + K_4^2)
+K_1^2 K_2^2 ((K_1\cdot K_4) + (K_2\cdot K_4) + K_4^2)\right], \cr
\beta_4 & =   {\pm   K_{12}^2 K_{41}^2 \rho   \over 4 P^2}.
}}
In particular, $|{\cal S}|=2$, and
the  two solutions $S_+,S_-$ are related by a change of sign of $\rho$, which appears in $\beta_4$.  Since there is a sum over the two solutions $S_+$ and $S_-$, the result for ${\hat a}_\alpha$ is seen to be rational, as it must be \BernKR.\foot{For one-, two-, and three-mass coefficients, $\rho$ turns out to be rational, so each solution is individually rational.}

We begin with the four-mass box integral, where the solution \ibciz\ can be used directly in the formula \besatway.\foot{The formula \ibciz\ is implicitly written in signature $(--++)$.  To perform a calculation with external momenta in Minkowski space, one must Wick-rotate $\epsilon_{\mu\nu\rho\sigma} \rightarrow i\epsilon_{\mu\nu\rho\sigma}$. }

For all other box integrals, we must work in the signature $(--++)$, as described in section 2.  As we have seen, the helicity configuration determines how to solve for the spinor components of the cut propagators.

It is worth noting that for all these cases, there are still just the two solutions \ibciz; in fact, a given solution determines whether the holomorphic or antiholomorphic spinors are proportional at each three-gluon vertex.  
This relation is illustrated in Figure 4, for each family of box functions with a three-gluon vertex.

\ifig\twosoln{Possible helicity assignments at three-gluon vertices, as derived from the two solutions  $S_+,S_-$ given in \nullshe.  The assignment $++-$ means that the solution dictates that the holomorphic spinors $\lambda$ are all proportional at this vertex; the assignment $--+$ means that the solution dictates that the antiholomorphic spinors $\lt$ are proportional.  The three helicities can then be distributed freely on the three legs at that vertex.
} {\epsfxsize=0.75\hsize\epsfbox{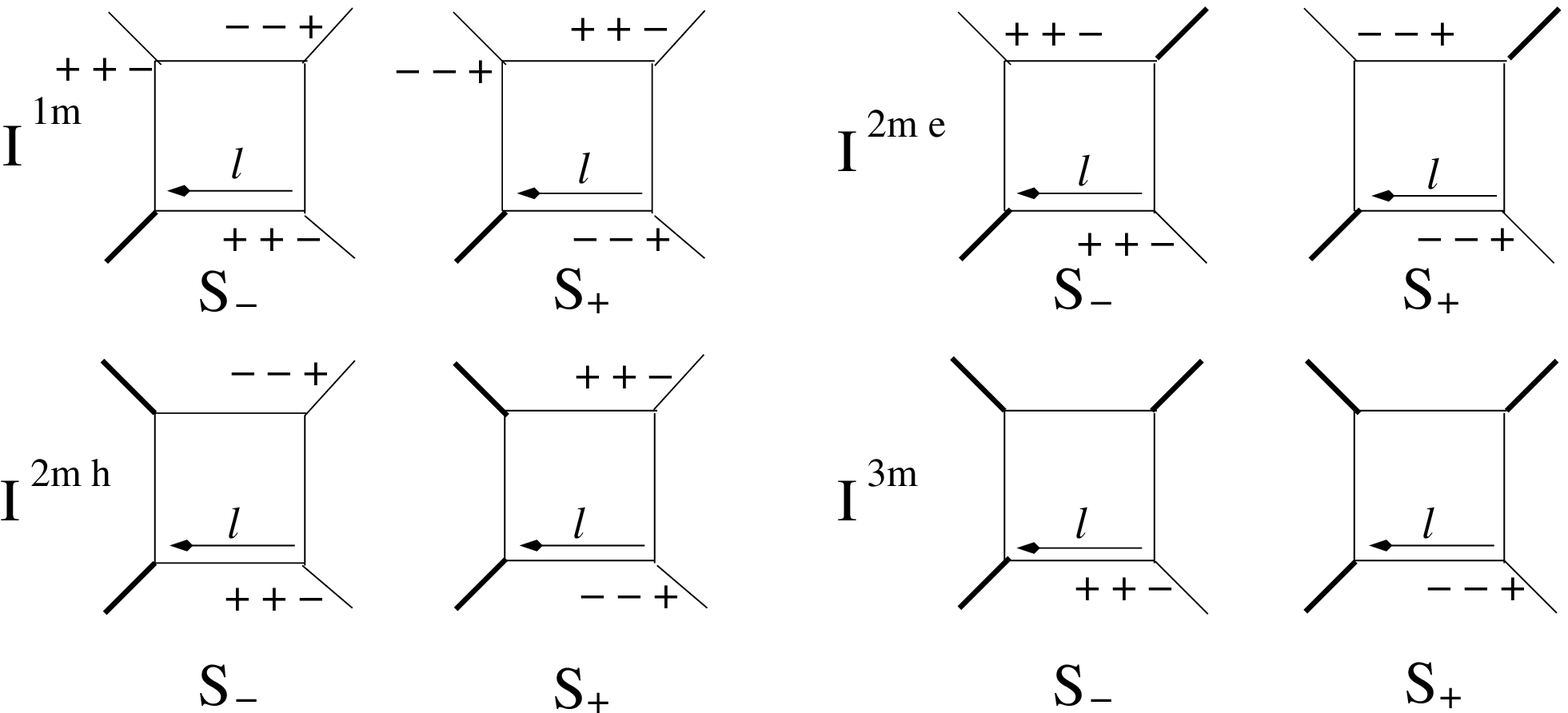}}

In practice, however, it is useful to solve for spinor components directly, rather than using \ibciz\ as written.  This is because the four tree-level amplitudes in \besatway\ can be easily computed from Parke-Taylor formulas \parke\ or MHV diagrams \refs{\CachazoKJ,\mhvdiag}, which involve spinor products.

In the three-mass example, we demonstrate the judicious solution of spinor components.
In the two-mass-easy example, we show that overall momentum conservation is an important constraint on the solutions of spinor components.  In the two-mass-hard example, we show that any pair of adjacent three-gluon corners must come with opposite helicity configurations (one each of $++-$ and $--+$), which is consistent with Figure 4.  This restriction figures into our one-mass example also, where we choose to compute a box integral whose quadruple cut has a next-to-MHV amplitude at one corner.

The four-mass coefficient presented here is new, but our examples in this section for three-,two-, and one-mass coefficients are all for a next-to-MHV seven-gluon amplitude that was computed in \BrittoNJ\ and \BernKY.  Our formulas reproduce the ones in those papers exactly.

In section 4, we will present several examples of infinite classes of coefficients for $n$-gluon next-to-MHV amplitudes.


\subsec{Four-Mass Box Integral Coefficients}

Let us illustrate our solution by computing the coefficients ${\hat f}_1,{\hat f}_2$ of the scalar box functions $I^{4m}_{8:2:2:2;1}$ and $I^{4m}_{8:2:2:2;2}$ for the helicity configuration $(1^-,2^-,3^-,4^-,5^+,6^+,7^+,8^+)$.  We can see immediately that ${\hat f}_1=0$, because any possible assignment of helicities for the internal lines, one of the four tree-level amplitudes in \besatway\ always vanishes.

The box diagram for $I^{4m}_{8:2:2:2;2}$ is shown in Figure 5.

\ifig\busyfig{The box function $I^{4m}_{8:2:2:2;2}$ for the helicity configuration $(1^-,2^-,3^-,4^-,5^+,6^+,7^+,8^+)$ with internal momenta $\ell_1,\ell_2,\ell_3,\ell_4$, and its associated quadruple cut.
} {\epsfxsize=0.75\hsize\epsfbox{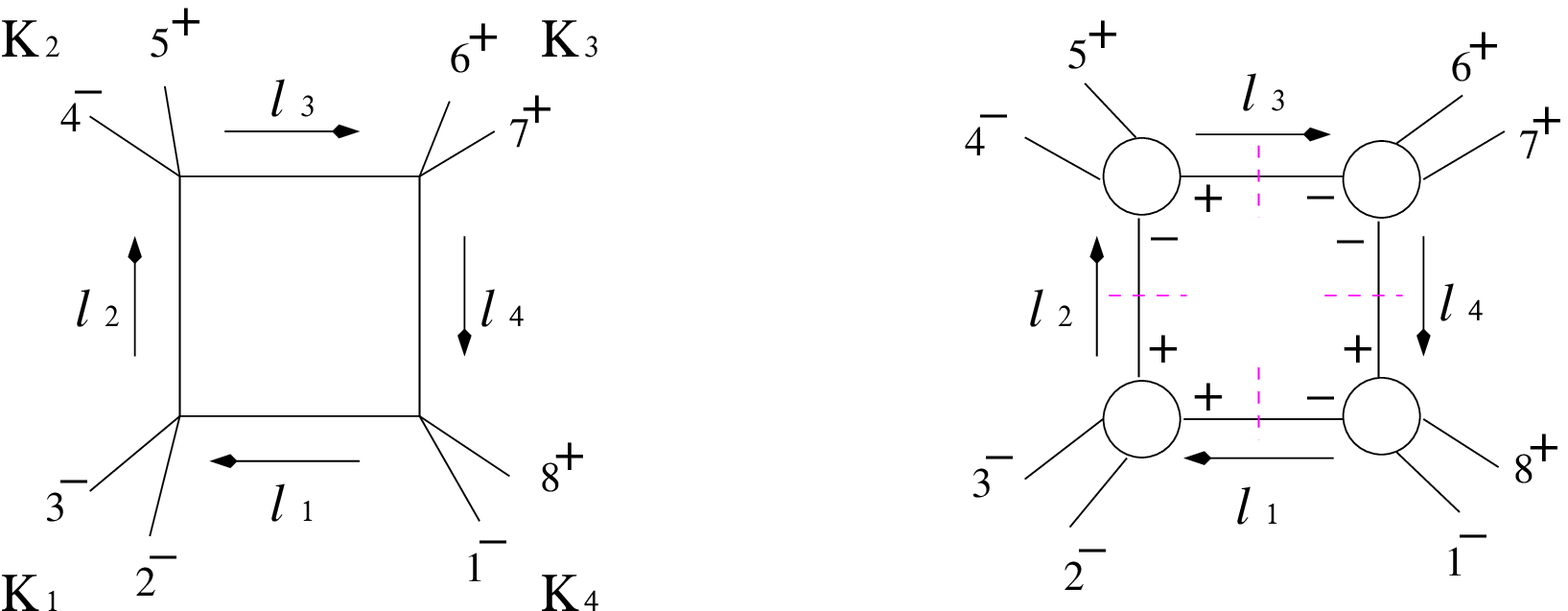}}

This helicity configuration allows only gluons to circulate in the loop.  Moreover, each vertex is MHV, so it is straightforward to write the coefficient \besatway\  in terms of Parke-Taylor formulas \parke.
\eqn\fourays{\eqalign{
{\hat f}_2
& = \half A^{\rm tree}((-\ell_4)^+,8^+,1^-,\ell_1^-) \times \cr
&~~~~~\times A^{\rm tree}((-\ell_1)^+,2^-,3^-,(-\ell_2)^+)
A^{\rm tree}(\ell_2^-,4^-,5^+,(-\ell_3)^+)
A^{\rm tree}(\ell_3^-,6^+,7^+,\ell_4^-) \cr
&= \half \sum_{S_+,S_-} {\vev{1~\ell_1}^3 \over \vev{\ell_1~\ell_4}\vev{\ell_4~8}\vev{8~1}}
{\vev{2~3}^3 \over \vev{3~\ell_2}\vev{\ell_2~\ell_1}\vev{\ell_1~2}}
{\vev{\ell_2~4}^3 \over \vev{4~5}\vev{5~\ell_3}\vev{\ell_3~\ell_2}}
{\vev{\ell_4~\ell_3}^3 \over \vev{\ell_3~6}\vev{6~7}\vev{7~\ell_4}} \cr
&= \half \sum_{S_+,S_-} {[6~7]^3 \vev{1|\slashed\ell_1~\slashed\ell_2|4}^3
\over \vev{8~1} [2~3] \vev{4~5}
\gb{5|\slashed\ell_3~\slashed\ell_4~\slashed\ell_1|2}
[3|\slashed\ell_2~\slashed\ell_3|6] \gb{8|\slashed\ell_4|7}}.
}}

Further information about four-mass box integrals, consistency checks for this coefficient, and twistor space structure may be found in the appendices.

\subsec{Three-Mass Example}

 Consider the three-mass box scalar integral for the seven-gluon next-to-MHV
amplitude $A_{7:1}(1^-,2^-,3^-,4^+,5^+,6^+,7^+)$ shown in
Figure 6.

\ifig\boxdef{A three-mass  scalar box integral for the
next-to-MHV amplitude $A_{7:1}(1^-,2^-,3^-,4^+,5^+,6^+,7^+)$.
} {\epsfxsize=0.75\hsize\epsfbox{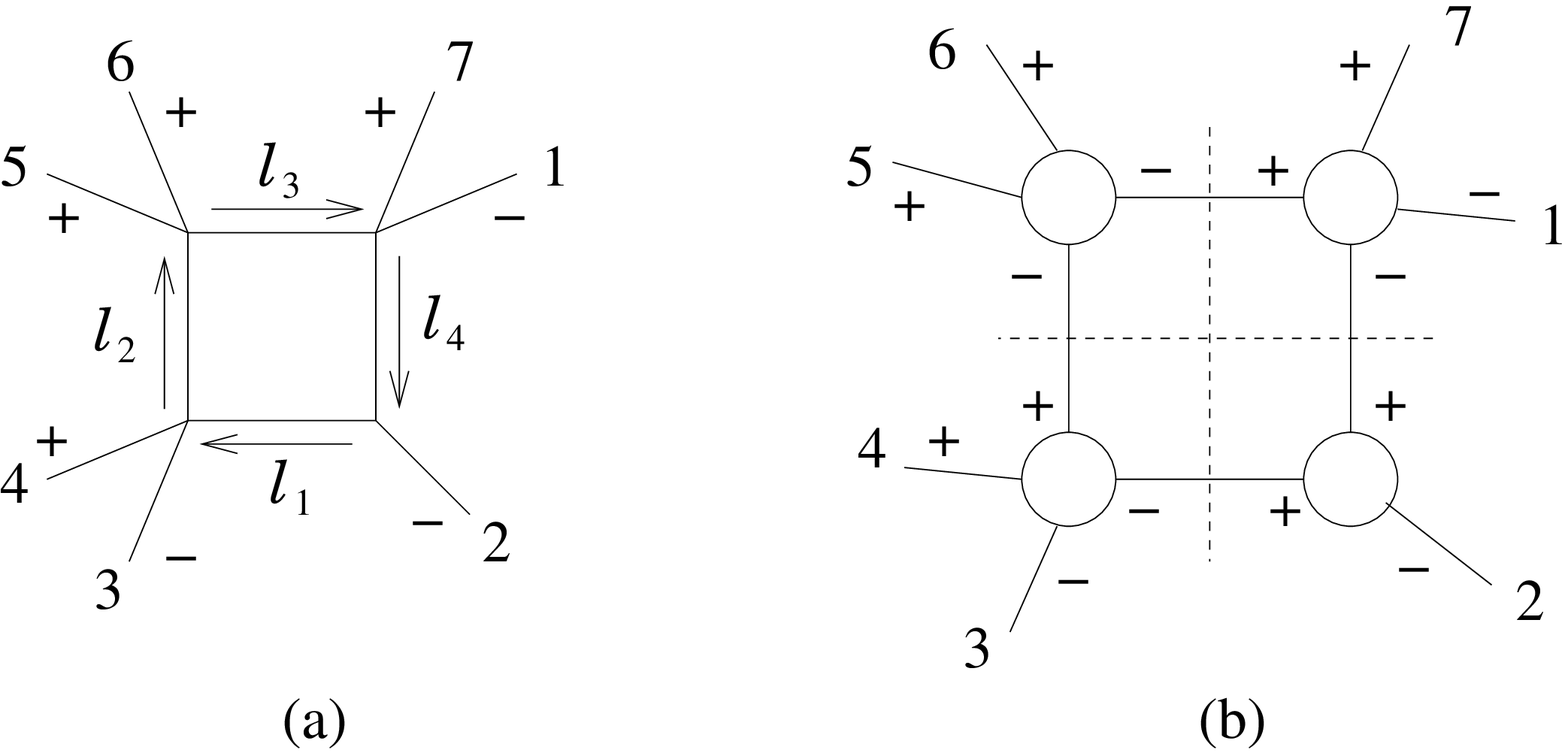}}

As explained in section 2, for the three-gluon amplitude with helicities $(++-)$,we should choose the $\lambda$'s to be proportional so that the amplitude does not vanish.
  We make this choice in solving the equations  $p_2\cdot \ell_1 =
p_2\cdot \ell_4 = \ell_1 \cdot \ell_4 = 0$, which can be written
as $\vev{ \ell_1 ~ 2}[\ell_1 ~ 2] =\vev{ \ell_4 ~ 2}[\ell_4 ~ 2]
=\vev{ \ell_1 ~ \ell_4}[\ell_1 ~ \ell_4] = 0 $.

We choose to write each four gluon amplitude in the generalized
unitarity cut diagram as a mostly minus MHV amplitude. This gives
\eqn\woau{ {\hat g}_3 = \half\left( { [\ell_1 ~\ell_4]^3\over
[\ell_1~2][2~\ell_4] } \right) \left( { [4~\ell_2]^3 \over
[\ell_2~\ell_1][\ell_1~3][3~4] } \right) \left( {[5~6]^3\over
[6~\ell_3] [\ell_3~\ell_2] [\ell_2 ~ 5] } \right) \left( {[\ell_3
~ 7]^3 \over [7~1] [1~ \ell_4 ] [\ell_4 ~ \ell_3 ] } \right).}
%

Now we explain how to solve $\lambda_{\ell_1}$ and $\tilde\lambda_{\ell_1}$
from light-cone conditions. First, using
$0=2\ell_1\cdot p_2 =\vev{2~\ell_1}[2~\ell_1]$, we get
$\vev{2~\ell_1}=0$ or $[2~\ell_1]=0$. To get a nonzero contribution, we choose to solve
$\vev{2~\ell_1}=0$, so that $\lambda_{\ell_1}=\a \lambda_2$. Now we can solve
for $\a$:
\eqn\kima{\eqalign{
 2\ell_1\cdot (p_3+p_4) & = (p_3+p_4)^2=-\a \gb{2|(3+4)|\ell_1};  \cr
\a & ={-(p_3+p_4)^2\over  \gb{2|(3+4)|\ell_1}}.} }
To solve for $\tilde\lambda_{\ell_1}$, we use
the equation $\ell_3^2 = (\ell_1- (p_3+p_4+p_5+p_6))^2 =0$, i.e.,
\eqn\kimaa{\eqalign{
 2\ell_1\cdot (p_3+p_4+p_5+p_6) & = (p_3+p_4+p_5+p_6)^2
=-\a \gb{2|(3+4+5+6)|\ell_1}  \cr
(p_3+p_4+p_5+p_6)^2 \gb{2|(3+4)|\ell_1} & =(p_3+p_4)^2 \gb{2|(3+4+5+6)|\ell_1}}}
Note that this is not
enough to fix the whole $\tilde\lambda_{\ell_1}$, but it fixes its
direction. This is good enough to compute \woau, since the
coefficient has degree zero in $\tilde\lambda_{\ell_1}$.

Then we do the same for $\ell_4$ and compute $\lambda_{\ell_4}$
and $\tilde\lambda_{\ell_4}$ up to a scale.

After this is done, we choose to write
\eqn\brass{ [\ell_2 ~ \bullet ] = { \gb{ 2 | \ell_2 | \bullet
}\over \vev{2 ~ \ell_2 }}, \qquad [ \ell_3 ~ \bullet ] = { \gb{ 2
| \ell_3 | \bullet }\over \vev{2 ~ \ell_3}}. }
where $\bullet$ represents any external gluon, $\ell_1$ or
$\ell_4$. It is easy to see that all denominators will drop out
from the fact that the coefficient has degree zero in $\ell_2$ and
$\ell_3$.

This gives for the coefficient \woau\ the following:
\eqn\fin{ {\hat g}_3 = \half\left( { [\ell_1 ~\ell_4]^3\over
[\ell_1~2][2~\ell_4] } \right) \left( { \gb{2| \ell_2 |4}^3 \over
\gb{2| \ell_2|\ell_1}[\ell_1~3][3~4] } \right) \left(
{[5~6]^3\over \gb{2|\ell_3 |6} \vev{2|\ell_3\cdot \ell_2 |2} \gb{2|
\ell_2 |5} } \right) \left( {\gb{2|\ell_3 | 7}^3 \over [7~1] [1~
\ell_4 ] \gb{2| \ell_3 | \ell_4 } } \right),  }
where we will substitute for $\ell_2$ and $\ell_3$ using
\eqn\plagio{
\ell_2 = \ell_1 - p_3 - p_4, \qquad \ell_3 = \ell_4+p_1+p_7.
}
The coefficient of the associated three-mass scalar box function
was computed by two different methods in \refs{\BrittoNJ,\BernKY}, and it is given by
\eqn\preww{\eqalign{
 & g_3 = -2{{\hat g}_3\over (p_3+p_4+p_5+p_6)^2(p_2+p_3+p_4)^2
- (p_3+p_4)^2(p_7+p_1)^2} \cr
&=  - { \vev{1~2}^3
\vev{2~3}^3 [5~6]^3 \over \vev{7~1} \vev{3~4} \gb{2 | 3+4| 5} \gb{ 2|
7+1|6} ( \vev{7~1} \gb{ 2| 3+4 |1} - t_{2}^{[3]} \vev{7~2}) (
t_{3}^{[4]}\vev{2~4} -\vev{3~4} \gb{ 2|7+1| 3})}.
}}

It is easy to check that the two formulas \fin,\preww\ give exactly the same
answer for ${\hat g}_3$.

\subsec{Two-Mass-Easy Example}

Consider the coefficient $c_4$ in the amplitude $A_{7:1}(1^-,2^-,3^-,4^+,5^+,6^+,7^+)$.  The box integral is shown in Figure 7.

\ifig\shale{Quadruple cut of a two-mass-easy scalar box integral for the amplitude $A_{7:1}(1^-,2^-,3^-,4^+,5^+,6^+,7^+)$.
} {\epsfxsize=0.75\hsize\epsfbox{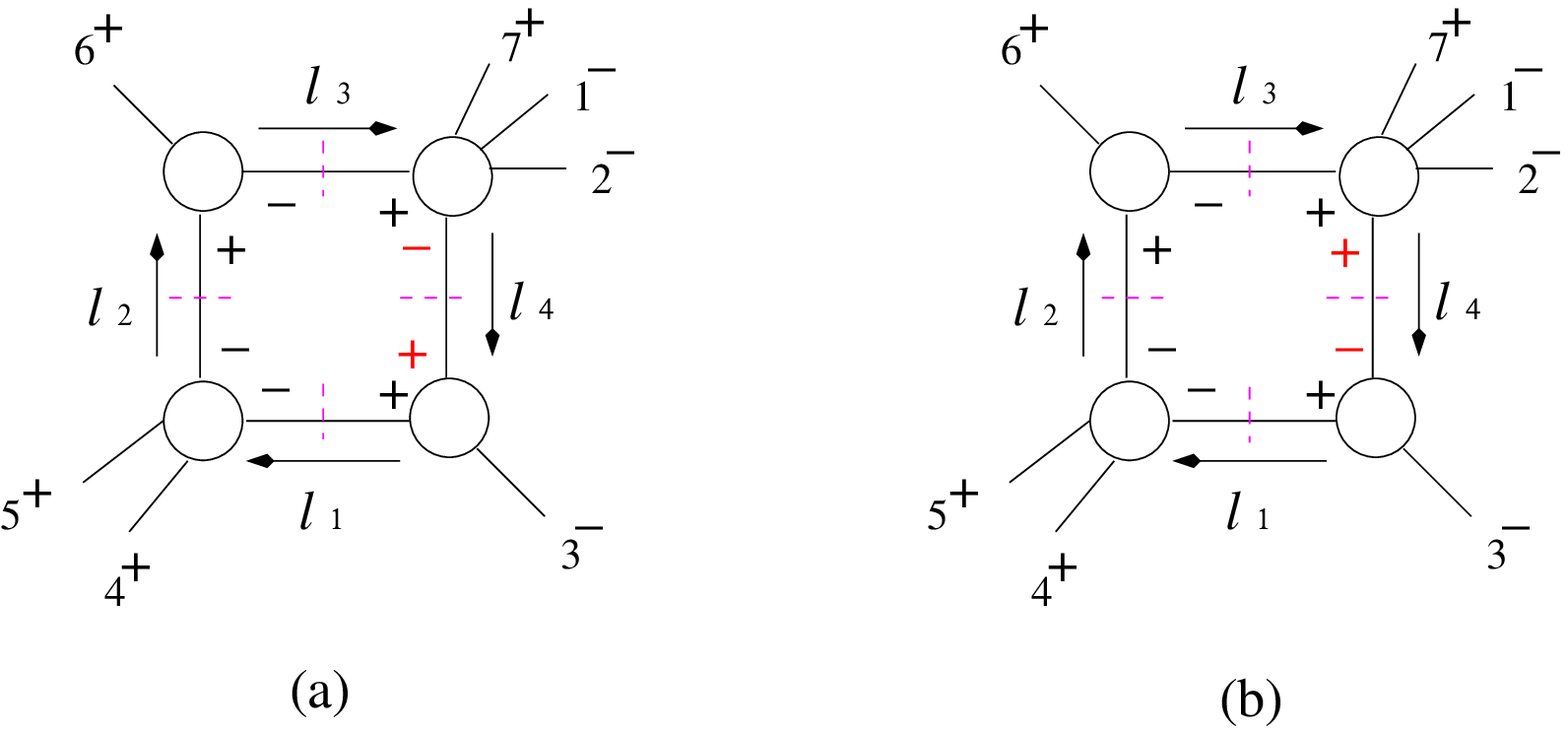}}

Notice that there are two possibly nonvanishing helicity assignments for the cut propagators.  Here it is possible to solve for the spinor components of $(\ell_1,\ell_2,\ell_3,\ell_4),$ up to a scale for each, for both helicity configurations.  However, the solution for configuration (b) found from this procedure fails to satisfy momentum conservation at the ``massive'' corners.  Therefore, there is only one true solution, namely configuration (a).  From it we find
\eqn\olivine{{\hat c}_4
=\half\left({\vev{\ell_2~\ell_1}^3 \over \vev{\ell_1~4}\vev{4~5}\vev{5~\ell_2}}\right)
\left({[\ell_2~6]^3 \over [6~\ell_3][\ell_3~\ell_2]}\right)
\left({[\ell_3~7]^3 \over [7~1] [1~2] [2~\ell_4] [\ell_4~\ell_3]}\right)
\left({[\ell_1~\ell_4]^3 \over [\ell_4~3][3~\ell_1]}\right).
}
Then
\eqn\mica{c_4=-2{{\hat c}_4 \over (p_4+p_5+p_6)^2 (p_3+p_4+p_5)^2
- (p_4+p_5)^2 (p_7+p_1+p_2)^2},
}
which agrees with the results given in \refs{\BrittoNJ,\BernKY}.

\subsec{Two-Mass-Hard Example}

Consider the coefficient $d_{3,2}$ in the amplitude $A_{7:1}(1^-,2^-,3^-,4^+,5^+,6^+,7^+)$.  The box integral is shown in Figure 8.

\ifig\garnet{Quadruple cut of a two-mass-hard scalar box integral for the amplitude $A_{7:1}(1^-,2^-,3^-,4^+,5^+,6^+,7^+)$.
} {\epsfxsize=0.75\hsize\epsfbox{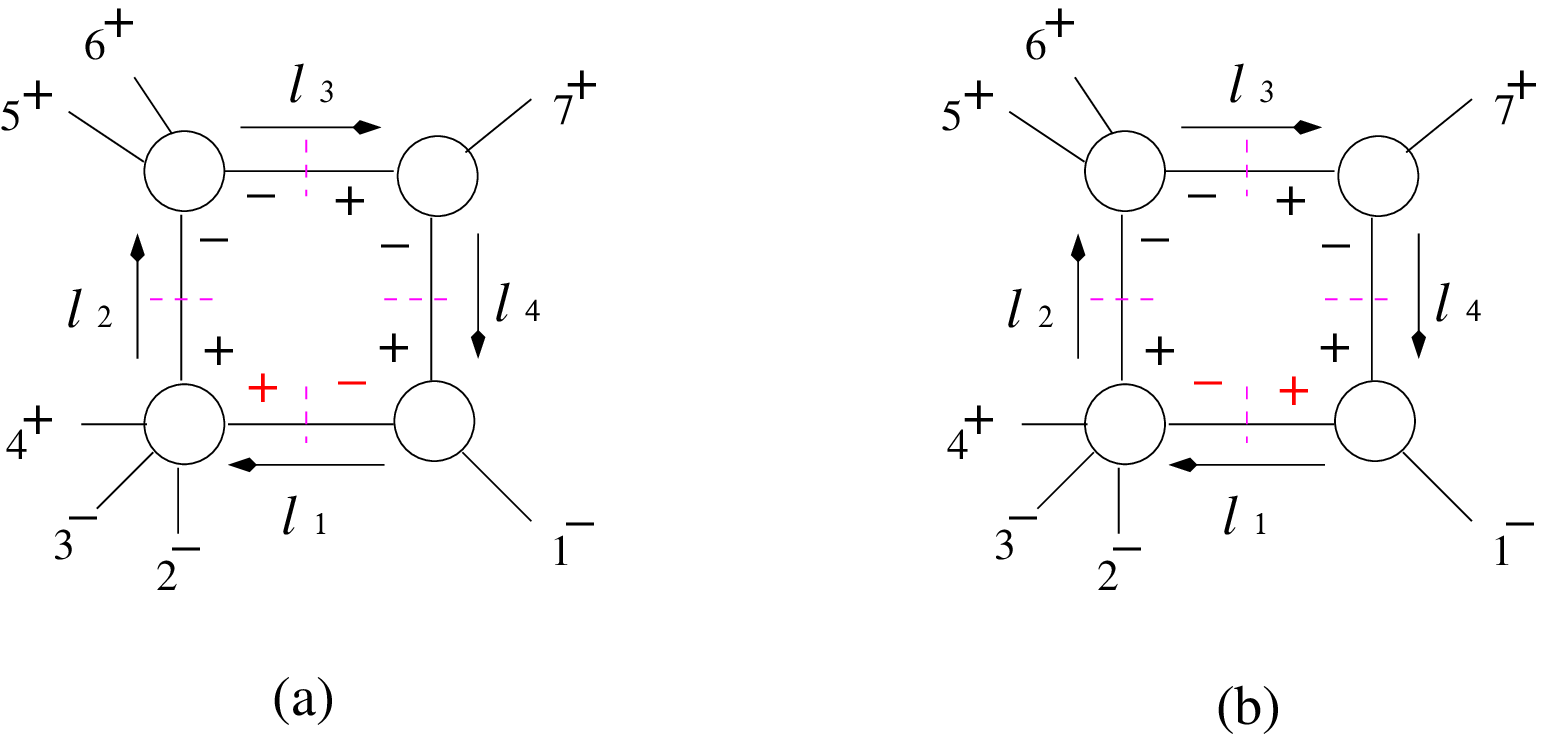}}

There are two possibly nonvanishing helicity assignments for the cut propagators.   For configuration (b), it is not
 possible to solve for the spinor components of $(\ell_1,\ell_2,\ell_3,\ell_4)$.  The reason is that from the $p_7$ corner, we derive the condition that $\lambda_{\ell_3},\lambda_{\ell_4}$ and $\lambda_7$ are all proportional, and similarly from the $p_1$ corner, we derive  that $\lambda_{\ell_4},\lambda_{\ell_1}$ and $\lambda_1$ are all proportional.  But this means that $\lambda_7$ is proportional to $\lambda_1$, which is not true of generic external momenta.
The lesson here is that adjacent three-gluon box corners must have opposite helicity types in order to give a nonvanishing product of amplitudes.

For the helicity configuration (a), there is no such obstacle to solving for  the spinor components of $(\ell_1,\ell_2,\ell_3,\ell_4)$ up to a scale, because the conditions are that  $\lambda_{\ell_3},\lambda_{\ell_4}$ and $\lambda_7$ are all proportional, but it is the antiholomorphic spinors  $\tilde\lambda_{\ell_4},\tilde\lambda_{\ell_1}$ and $\tilde\lambda_1$ that are proportional for the other corner.  Once the solution is obtained, the coefficient ${\hat d}_{3,2}$ is given by
\eqn\obsidian{{\hat d}_{3,2}=
\half \left({\vev{\ell_1~1}^3 \over \vev{1~\ell_4}\vev{\ell_4~\ell_1}}\right)
\left({[\ell_3~7]^3 \over [\ell_3~\ell_4][\ell_4~\ell_7]}\right)
\left({\vev{\ell_3~\ell_2}^3 \over \vev{\ell_2~5}\vev{5~6}\vev{6~\ell_3}}
\right)
\left({\vev{2~3}^3 \over \vev{3~4}\vev{4~\ell_2}\vev{\ell_2~\ell_1}\vev{\ell_1~2}}\right).
}
Then
\eqn\hematite{d_{3,2}=-2{{\hat d}_{3,2} \over (p_5+p_6+p_7)^2 (p_7+p_1)^2},
}
which agrees with the results of \refs{\BrittoNJ,\BernKY}.

\subsec{One-Mass Example}

Even for the one-mass box functions, where the internal momenta might seem to be overconstrained, we can solve for the coefficients using generalized unitarity.
Consider the coefficient $b_2$ in the amplitude $A_{7:1}(1^-,2^-,3^-,4^+,5^+,6^+,7^+)$.  The box integral is shown in Figure 9.  We have applied the lesson learned in the previous subsection, that  adjacent three-gluon box corners must have opposite helicity types in order to give a nonvanishing product of amplitudes.
Thus only the two helicity configurations shown in the diagram might contribute to the coefficient.

\ifig\quartz{Quadruple cut of a one-mass scalar box integral for the
amplitude $A_{7:1}(1^-,2^-,3^-,4^+,5^+,6^+,7^+)$.
} {\epsfxsize=0.80\hsize\epsfbox{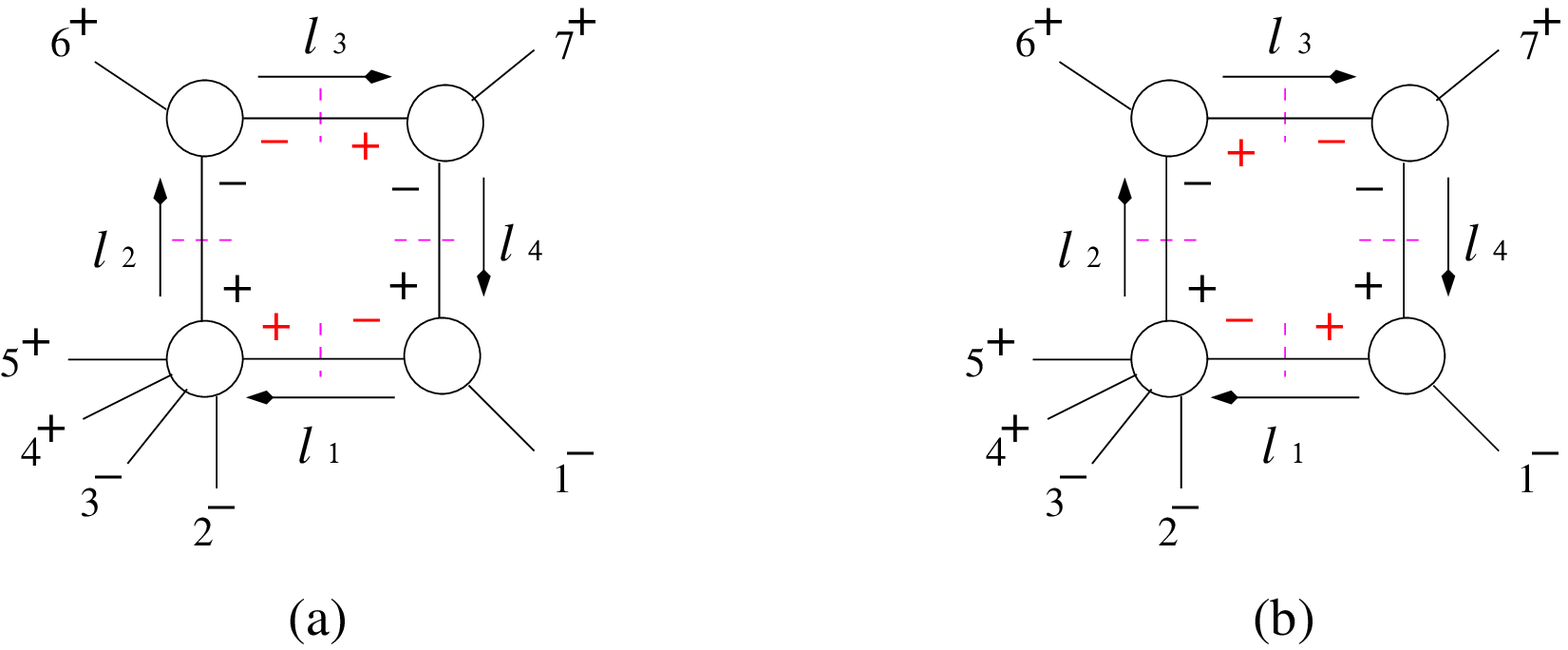}}

For the configuration (a), we solve for the spinor components of $(\ell_1,\ell_2,\ell_3,\ell_4)$ while requiring that
\eqn\trilobite{
[\ell_2~6]=[\ell_3~6]=[\ell_2~\ell_3]=0, \quad
\vev{\ell_3~7}=\vev{\ell_4~7}=\vev{\ell_3~\ell_4}, \quad
[\ell_4~1]=[\ell_1~1]=[\ell_4~\ell_1]=0.
}
Then we substitute this solution into the product of tree amplitudes:
\eqn\pyrrhite{
{\hat b}_2^{(a)} = 
\half\left({\vev{2~3}^3 \over \vev{3~4}\vev{4~5}\vev{5~\ell_2}\vev{\ell_2~\ell_1}\vev{\ell_1~2}}\right)
\left({\vev{\ell_2~\ell_3}^3 \over \vev{\ell_2~6}\vev{6~\ell_3}}\right)
\left({[\ell_3~7]^3 \over [7~\ell_4] [\ell_4~\ell_3]}\right)
\left({\vev{1~\ell_1}^3 \over \vev{\ell_1~\ell_4}\vev{\ell_4~1}}\right).
}

For the configuration (b), we solve for the spinor components of $(\ell_1,\ell_2,\ell_3,\ell_4)$ while requiring that
\eqn\pumice{
\vev{\ell_2~6}=\vev{\ell_3~6}=\vev{\ell_2~\ell_3}=0, \quad
[\ell_3~7]=[\ell_4~7]=[\ell_3~\ell_4], \quad
\vev{\ell_4~1}=\vev{\ell_1~1}=\vev{\ell_4~\ell_1}=0.
}
We use this solution in the product of tree amplitudes, which now includes an NMHV amplitude \refs{\mangpxu,\mangparke}:
\eqn\mpxsix{\eqalign{A_6^{\rm tree}((-\ell_1)^-, 2^-,3^-,& 4^+,5^+,\ell_2^+)  = 
 \left[ {\beta^2 \over t_{5\ell_2 (-\ell_1)}s_{5 \ell_2 }s_{\ell_2 (-\ell_1)}s_{23}s_{34}} \right. \cr
 & +\left.
{\gamma^2 \over t_{\ell_2 (-\ell_1) 2}s_{\ell_2 (-\ell_1)}s_{(-\ell_1)2}s_{3 4}s_{45}}+
{\beta \gamma
t_{45 \ell_2 } \over s_{45}s_{5\ell_2}s_{\ell_2 (-\ell_1)}s_{(-\ell_1) 2}s_{23}s_{34}} \right], \cr
\beta &= [5~\ell_2]\vev{2~3}\gb{\ell_1|5+\ell_2|4}, \cr \gamma &=
[4~5]\vev{\ell_1~2}\gb{3|4+5|\ell_2}, \cr
s_{ij}&=\vev{i~j}[i~j], \cr
t_{ijk}&=\vev{i~j}[i~j]+\vev{i~k}[i~k]+\vev{j~k}[j~k]
.}}
The contribution of this configuration to the coefficient is
\eqn\basalt{
{\hat b}_2^{(b)}=\half A_6^{\rm tree}((-\ell_1)^-,2^-,3^-,4^+,5^+,\ell_2^+) \times
\left({[6~\ell_3]^3 \over [\ell_3~\ell_2][\ell_2~6]}\right)
\left({\vev{\ell_4~\ell_3}^3 \over \vev{\ell_3~7}\vev{7~\ell_4}}\right)
\left({[\ell_1~\ell_4]^3 \over [\ell_4~1][1~\ell_1]}\right).
}
Then
\eqn\beryl{b_2=-2{{\hat b}_2^{(a)} +{\hat b}_2^{(b)} \over (p_6+p_7)^2 (p_7+p_1)^2,}
}
which agrees with the results of \refs{\BrittoNJ,\BernKY}.
We note that in \BrittoNJ, this coefficient $b_2$ was given in terms of a seven-gluon tree amplitude and eight other coefficients (using a relation derived from the infrared singular behavior of a certain unitarity cut).  Here, the quadruple cut allows us to obtain an explicit formula for $b_2$ directly.

One can easily obtain $b_2$ for the $n$-gluon next-to-MHV amplitude $(1^-,2^-,3^-,4^+,...,n^+)$ by substituting the tree amplitude $ A_{n-1}^{\rm tree}((-\ell_1)^-,2^-,3^-,4^+,...,(n-2)^+,\ell_2^+)$ for $A_6^{\rm tree}$ in \basalt.  Compact expressions for that amplitude have been given in \refs{\CachazoKJ,\KosowerYZ}.

\newsec{All-Multiplicity Examples for NMHV Amplitudes}

In this section we present some examples of using quadruple cuts to compute some classes of coefficients for $n$-gluon next-to-MHV amplitudes.  Here we substitute the actual solutions for the cut propagators, so that our final formulas are given in terms of the external momenta only.

\subsec{All-Multiplicity Examples of Three-Mass Coefficients}

In this subsection we present three classes of three-mass coefficients for next-to-MHV $n$-gluon amplitudes.  The first two are depicted in Figure 10.

\ifig\obol{Two families of three-mass scalar box integrals in an NMHV configuration.
} {\epsfxsize=0.70\hsize\epsfbox{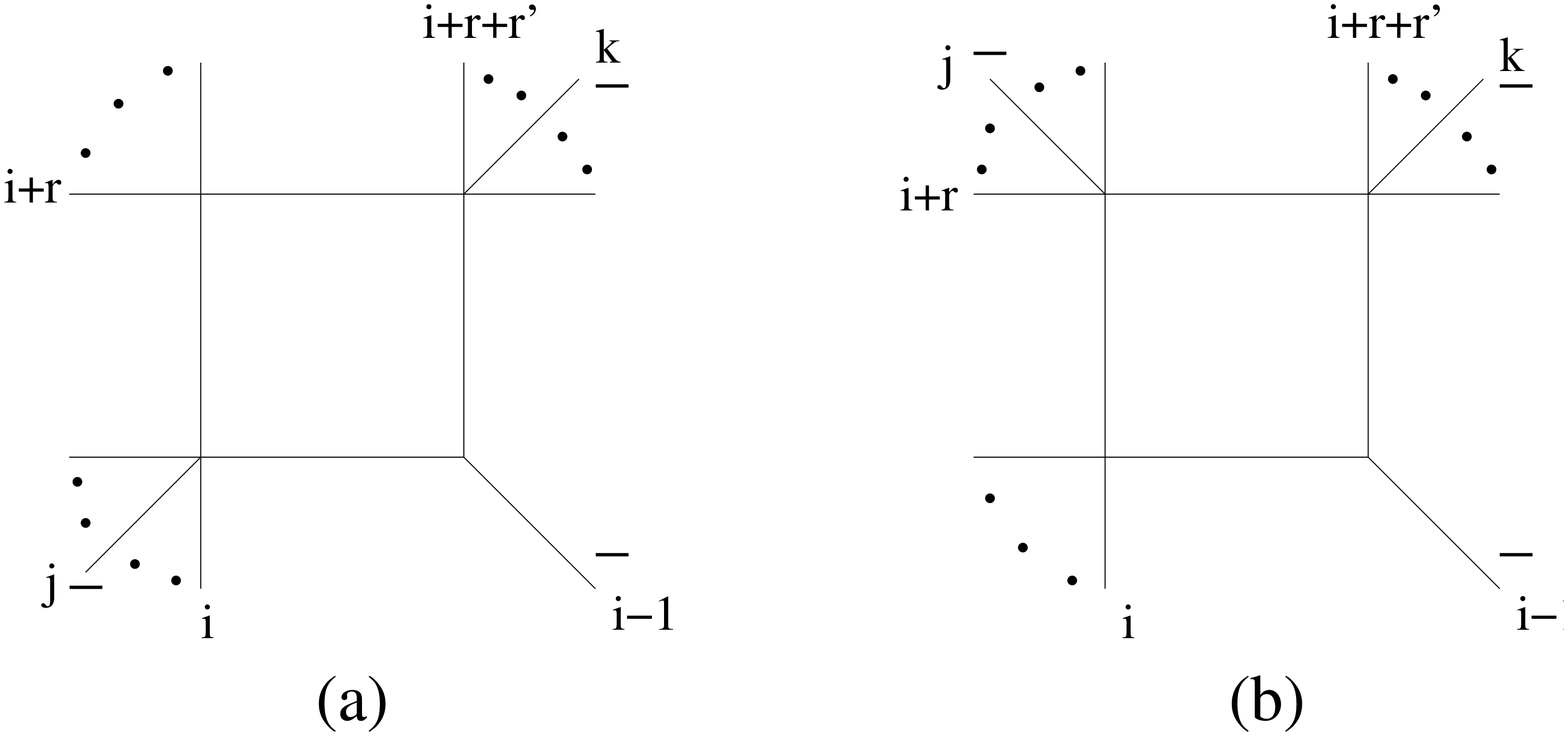}}

Consider first the three-mass scalar box integral coefficients ${\hat g}_{n:r:r'; i}$,
where the
three negative helicities are $j, k, i-1$, and $j$ is in the vertex of
$(i,...,i+r-1)$ and $k$ is in the vertex of $(i+r+r',...,i-2)$.  See Figure 10(a).
 Using \besatway, we can write
down
\eqn\shekel{\eqalign{
& {\hat g}_{n:r:r'; i} ={\vev{i-2~i-1} \vev{i-1~i}\vev{i+r-1~i+r}
\vev{i+r+r'-1~ i+r+r'} \over 2 \prod_{s=1}^{n} \vev{ s~s+1}}
{[\ell_1 ~\ell_4]^3 \over [\ell_4~i-1][i-1~\ell_1]} \cr
& \times { \vev{ \ell_1~ j}^4 \over \vev{\ell_2~ \ell_1}\vev{\ell_1~ i}
\vev{i+r-1~\ell_2}} {\vev{ \ell_3~
\ell_2}^3 \over \vev{\ell_2~i+r}
\vev{i+r+r'-1~\ell_3}}
{\vev{k~\ell_4}^4 \over
\vev{\ell_3~i+r+r'}
\vev{i-2~\ell_4} \vev{\ell_4~\ell_3}}.}}

We simplify the above formula by noticing that
\eqn\dinar{\eqalign{
\vev{ p| K_2\cdot K_1| i-1} &= \vev{p~\ell_3}[\ell_3~\ell_2]\vev{\ell_2~i-1} \cr\vev{ p| K_2\cdot K_3| i-1} &= \vev{p~\ell_2}[\ell_2~\ell_3]\vev{\ell_3~i-1}
}}
and $\lambda_{\ell_1}=\a \lambda_{i-1},~\lambda_{\ell_4}=\b
\lambda_{i-1}$,
so
\eqn\tuppence{\eqalign{ {\hat g}_{n:r:r'; i} &=
{\vev{i-2~i-1} \vev{i-1~i}\vev{i+r-1~i+r}
\vev{i+r+r'-1~ i+r+r'} \over 2 \prod_{s=1}^{n} \vev{ s~s+1}}
{\vev{i-1~ j}^4 \vev{k~i-1}^4 \over \vev{i-1~ i}
\vev{i-2~i-1}}\cr
& \times { (-K_2^2)^3 \over
\vev{ i+r-1| K_2\cdot K_3|i-1 }\vev{ i+r| K_2\cdot K_3|i-1 }}\cr
& \times {1 \over
\vev{i+r+r'-1| K_2\cdot K_1| i-1}\vev{i+r+r'| K_2\cdot K_1|
i-1}}
\cr
& \times  {[\ell_1 ~\ell_4]^3 \over  [\ell_4~i-1][i-1~\ell_1]}
\left(\a^2 \b^2 \vev{i-1|\ell_3\cdot \ell_2|i-1}\right).
}}
We will show below that the last line is simply $\left( K_{41}^2 K_{12}^2-K_1^2 K_3^2\right)$.
 Thus we get immediately
the
coefficients
\eqn\drachma{\eqalign{& g_{n:r:r';i}  =
{\vev{i-1~ j}^4
\vev{k~i-1}^4\vev{i+r-1~i+r}
\vev{i+r+r'-1~ i+r+r'} (K_2^2)^3 \over \prod_{s=1}^{n} \vev{ s~s+1}\vev{ i+r-1| K_2\cdot K_3|i-1 }} \cr
& \times  {1 \over \vev{ i+r| K_2\cdot K_3|i-1 }
\vev{i+r+r'-1| K_2\cdot K_1| i-1}\vev{i+r+r'| K_2\cdot K_1|
i-1}}.}}
It can be shown that if we put $j=3, i-1=2, k=1$ we reproduce the all-multiplicity NMHV
coefficients given in \BernKY. Also, our general formula matches the results of \BernKY\ for seven
gluons, i.e., the coefficient $c_{146}$ of helicity assignment
$(1^-,2^-,3^+,4^-,5^+,6^+,7^+)$ and $c_{257}$ of helicity assignment
$(1^-,2^+,3^-,4^+,5^-,6^+,7^+)$.

Now we prove the identity involving the last line of \drachma, namely that
\eqn\bigbux{ {[\ell_1 ~\ell_4]^3 \over [\ell_4~i-1][i-1~\ell_1]}
\left(\a^2 \b^2 \vev{i-1|\ell_3\cdot \ell_2|i-1}\right)=
K_{41}^2 K_{12}^2-K_1^2 K_3^2.}
To do this, we need to
notice
that
\eqn\pfennig{
\tilde\lambda_{\ell_4}={1\over\b}\tilde\lambda_{i-1}+{\a \over \b}\tilde
\lambda_{\ell_1},
}
which can be obtained by noticing that ${\ell_4}={\ell_1}+p_{i-1}$
while $p_{i-1}=\lambda_{i-1}\tilde\lambda_{i-1}$,
${\ell_1}= \a \lambda_{i-1}\tilde\lambda_{\ell_1}$,
${\ell_4}= \b \lambda_{i-1}\tilde\lambda_{\ell_4}$. Using this we can
easily show that
\eqn\khoum{\eqalign{
& \left( K_{41}^2 K_{34}^2-K_1^2 K_3^2\right) =
K_1^2(2p_{i-1}\cdot K_3)+ K_3^2 (2p_{i-1}\cdot K_1)
+(2p_{i-1}\cdot K_3)(2p_{i-1}\cdot K_1)\cr
& =  (-)\vev{\ell_1~\ell_2}[\ell_1~\ell_2]
\vev{i-1~\ell_3}[i-1~\ell_3]
+(-)\vev{\ell_4~\ell_3}[\ell_4~\ell_3](-)\vev{i-1~\ell_2}[i-1~\ell_2]\cr &
 ~~~~+
\vev{i-1~\ell_3}[i-1~\ell_3](-)\vev{i-1~\ell_2}[i-1~\ell_2]\cr
& =  -\a \vev{i-1~\ell_2}[\ell_1~\ell_2]
\vev{i-1~\ell_3}[i-1~\ell_3]
+\b\vev{i-1~\ell_3}[\ell_4~\ell_3]\vev{i-1~\ell_2}[i-1~\ell_2]\cr &
~~~~-
\vev{i-1~\ell_3}[i-1~\ell_3]\vev{i-1~\ell_2}[i-1~\ell_2]\cr
& =  -\a \vev{i-1~\ell_2}
\vev{i-1~\ell_3}[\ell_1~i-1][\ell_2~\ell_3]= \a 
[i-1~\ell_1]\vev{i-1~\ell_3}[\ell_3~\ell_2]\vev{i-1~\ell_2}
}}
and
\eqn\picayune{\eqalign{
 & {[\ell_1 ~\ell_4]^3 \over [\ell_4~i-1][i-1~\ell_1]}
\left(\a^2 \b^2 \vev{i-1|\ell_3\cdot \ell_2|i-1}\right)\cr
& =  {{1 \over \b^3}[\ell_1~i-1]^3 \over
{\a \over\b}[\ell_1~i-1][i-1~\ell_1]}
\a^2 \b^2\vev{i-1~\ell_3}[\ell_3~\ell_2]\vev{\ell_2~i-1}\cr
& =  \a [i-1~\ell_1]\vev{i-1~\ell_3}[\ell_3~\ell_2]\vev{\ell_2~i-1}.
}}


By similar calculations, we have found another
series of coefficients of $F^{3m}$.
where the three negative helicities $j,k,i-1$ are at
the vertices $(i+r,...,i+r+r'-1)$, $(i+r+r',...,i-2)$
and massless leg $(i-1)$ respectively. The coefficient is given by
\eqn\drachmatwo{\eqalign{ g_{n:r:r';i}  = &
{\vev{k~~i-1}^4 \vev{j|K_2\cdot K_1|i-1}^4\vev{i+r-1~~i+r}
\vev{i+r+r'-1~~ i+r+r'}  \over (-K_2^2)\prod_{s=1}^{n} \vev{
s~s+1}}
\cr
&   \times {1\over
\vev{ i+r-1| K_2\cdot K_3|i-1 }\vev{ i+r| K_2\cdot K_3|i-1 }}
\cr
&   \times {1\over
\vev{i+r+r'-1| K_2\cdot K_1| i-1}\vev{i+r+r'| K_2\cdot K_1|
i-1}}.
}}

Now we offer an example of box integrals whose quadruple cut involves fermions and scalars, in addition to gluons, circulating in the loop.  
\ifig\obol{Another family of three-mass scalar box integrals in an NMHV configuration.
} {\epsfxsize=0.70\hsize\epsfbox{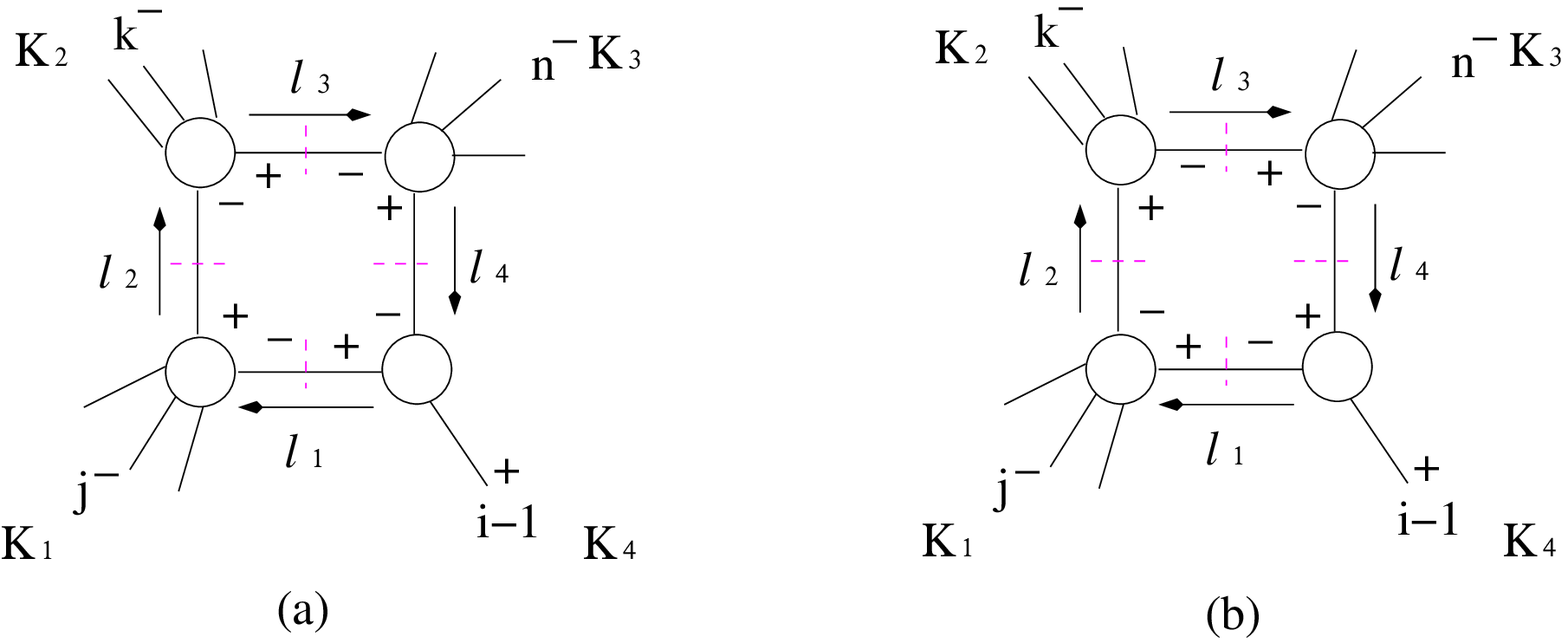}}
In this case, the three negative helicities $j,k,n$ are at 
vertex $K_1,K_2$ and $K_3$ respectively. There are two
possible helicity configurations for loop momentum.
See Figure 11.
 Using Ward identities, one can prove the following relations between amplitudes involving gluons and amplitudes involving fermions and scalars \refs{\grisaru,\mangparke}.
\eqn\MHVF{\eqalign{ A(
F_1^-, g_2^+,.., g_j^-,..., F_n^+) & =  { \vev{ j ~ n}\over \vev{j
~  1}} A^{\rm MHV}( g_1^-, g_2^+,..., g_j^-,..., g_n^+), \cr A( S_1^-,
g_2^+,.., g_j^-,..., S_n^+) & = { \vev{ j  ~ n}^2\over \vev{j ~
1}^2} A^{\rm MHV}( g_1^-, g_2^+,..., g_j^-,..., g_n^+), }}
we can write down the sum of configurations (a) and (b) as
\eqn\AandBone{\eqalign{&
{\vev{i-2~i-1} \vev{i-1~i}\vev{i+r-1~i+r}
\vev{i+r+r'-1~ i+r+r'}  \over 2 \prod_{s=1}^{n} \vev{ s~s+1}}
\cr & \times
{\vev{\ell_1~j}^2\vev{\ell_2~j}^2 \vev{\ell_2~k}^2
\vev{\ell_3~k}^2 \vev{\ell_3~n}^2\vev{\ell_4~n}^2
\over \vev{\ell_1~i}
\vev{i+r-1~\ell_2}\vev{\ell_2~\ell_1}\vev{\ell_2~i+r}
\vev{i+r+r'-1~\ell_3}\vev{\ell_3~\ell_2}\vev{\ell_3~i+r+r'}
\vev{i-2~\ell_4}\vev{\ell_4~\ell_3}} \cr
&  \times  {[\ell_4~i-1]^2 [\ell_1~i-1]^2 
\over [i-1~\ell_1][\ell_1~\ell_4]
[\ell_4~i-1]} \times A^{-2}\left( A-1\right)^4,
}} 
where
\eqn\Afactor{\eqalign{ A & =
 { \vev{j~\ell_1}\vev{k~\ell_2}\vev{n~\ell_3}[i-1~\ell_1]
\over \vev{j~\ell_2}\vev{k~\ell_3}\vev{n~\ell_4}[i-1~\ell_4]}\cr
&={ \vev{k|K_2\cdot K_3|i-1}\vev{n|K_2\cdot K_1|i-1}
 \over  \vev{j|K_2\cdot K_3|i-1}\vev{k|K_2\cdot K_1|i-1}}
\times {\gb{j|\ell_1|i-1}\over \gb{n|\ell_1|i-1}}.
}}
After simplifying the above expression similarly to the previous examples,
we obtain
\eqn\gfirst{\eqalign{&  g_{n:r:r':i}  = \cr &
{ \left( \vev{k|K_2\cdot K_3|i-1}\vev{n|K_2\cdot K_1|i-1}
\vev{j~i-1}- \vev{j|K_2\cdot K_3|i-1}
\vev{k|K_2\cdot K_1|i-1}\vev{n~i-1}\right)^4
\over\vev{ i+r-1| K_2\cdot K_3|i-1 }\vev{ i+r| K_2\cdot K_3|i-1 }
\vev{i+r+r'-1| K_2\cdot K_1| i-1}}
\cr
&  \times { \vev{i+r-1~i+r}
\vev{i+r+r'-1~ i+r+r'} \over \vev{i+r+r'| K_2\cdot K_1| 
i-1} K_2^2 \vev{i-1|K_2\cdot 
K_3|i-1}^4
\prod_{s=1}^{n} \vev{ s~s+1}}.
}}
To simplify further, we use the following result:
\eqn\middle{\eqalign{ &
\vev{k|K_2\cdot K_3|i-1}\vev{n|K_2\cdot K_1|i-1}
\vev{j~i-1}- \vev{j|K_2\cdot K_3|i-1}
\vev{k|K_2\cdot K_1|i-1}\vev{n~i-1}\cr
& =-\vev{i-1|K_2\cdot K_3|i-1} \left(
\vev{k|K_2\cdot K_3|n}\vev{j~i-1}+\vev{k|K_2\cdot K_1|j} 
\vev{n~i-1}\right. \cr &~~~~~~~~~~~~~~~~~~~~~~~~~~~~~~~~~~~~~~~~~~~~~
\left.-\vev{n~i-1}\vev{j~i-1}\gb{k|K_2|i-1}\right).
}}
We arrive at the final expression,
\eqn\gsecond{\eqalign{  g_{n:r:r':i} & =
{ \left( 
\vev{k|K_2\cdot K_3|n}\vev{j~i-1}+\vev{k|K_2\cdot (K_1+p_{i-1})|j} 
\vev{n~i-1}\right)^4
\over\vev{ i+r-1| K_2\cdot K_3|i-1 }\vev{ i+r| K_2\cdot K_3|i-1 }
\vev{i+r+r'-1| K_2\cdot K_1| i-1}}
\cr
&  \times { \vev{i+r-1~i+r}
\vev{i+r+r'-1~ i+r+r'} \over  \vev{i+r+r'| K_2\cdot K_1| 
i-1} K_2^2 
\prod_{s=1}^{n} \vev{ s~s+1}}.
}}
We have checked that this expression reproduces the
coefficient $c_{135}$
of seven gluons in the  helicity configuration $(1^-,2^-,3^+,4^-,5^+,6^+,7^+)$
as given in \BernKY.

\subsec{An All-Multiplicity Example of a Two-Mass-Hard Coefficient}

Now we consider a series of $F^{2m~h}$ shown in Figure 12, where
both massless legs $(i-1)$ and $(i-2)$ are negative helicities.

\ifig\copper{Quadruple cuts of a family of two-mass-hard scalar box integrals in an NMHV configuration.
} {\epsfxsize=0.85\hsize\epsfbox{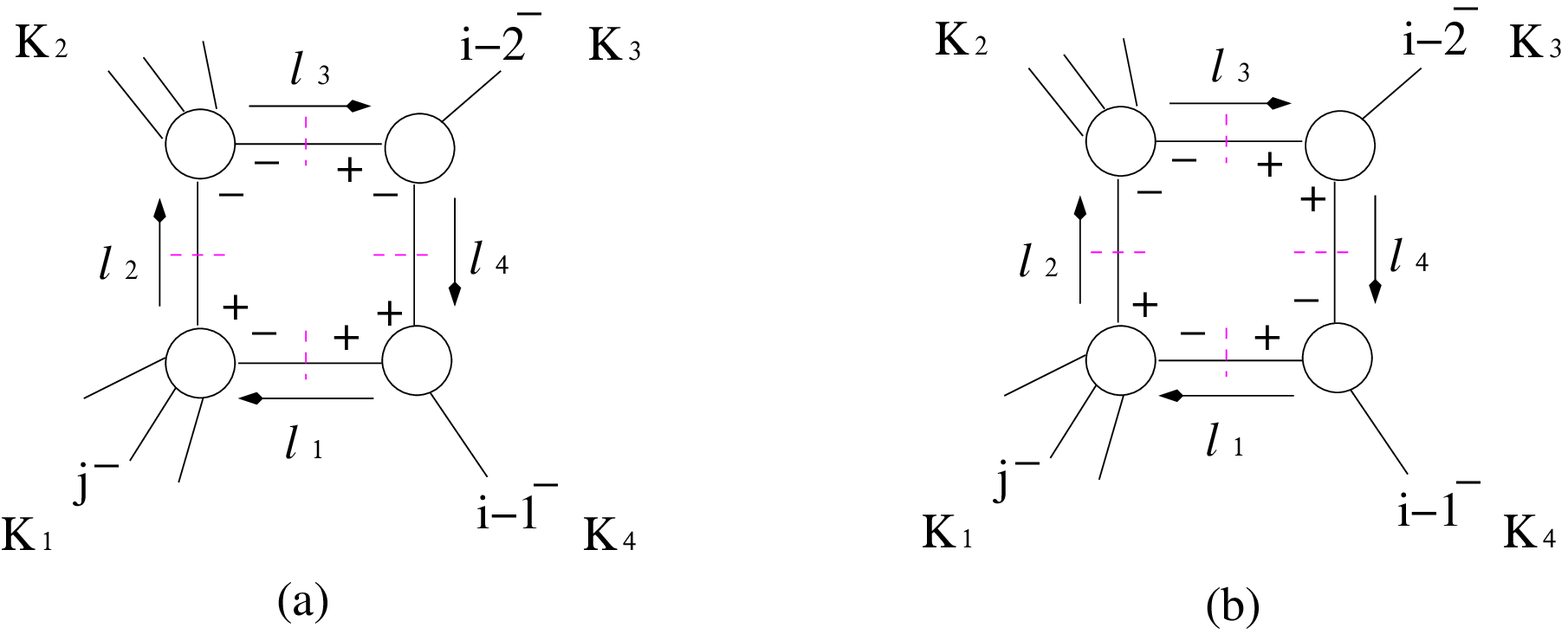}}
The third
negative helicity $j$ is at the vertex of $(i,...,i+r-1)$. For this
case there are two contributions. For part (a) we have
\eqn\parta{\eqalign{& I_a  ={\vev{i-3~i-2}\vev{i-2~i-1}\vev{i-1~i}\vev{i+r-1~i+r}
\over 2 \prod_{s=1}^{n} \vev{s~s+1}}
\cr & \times 
{[\ell_1 ~\ell_4]^3 \over
[\ell_4~i-1][i-1~\ell_1]}
 { \vev{\ell_1~j}^4\over \vev{\ell_1~i}\vev{i+r-1~\ell_2}
\vev{\ell_2~\ell_1 }} {\vev{\ell_3~\ell_2}^3 \over
\vev{\ell_2~i+r}\vev{i-3~\ell_3}} {\vev{i-2~\ell_4}^3
\over \vev{\ell_3~i-2}\vev{\ell_4~\ell_3}}.
}}
 Using similar
manipulations, with
$\lambda_{\ell_1}=\a \lambda_{i-1}$, $\lambda_{\ell_4}=\b \lambda_{i-1}$
and the identity \bigbux, which here takes the form
\eqn\identa{\eqalign{
{[\ell_1 ~\ell_4]^3 \over  [\ell_4~i-1][i-1~\ell_1]}
\times \a^2 \b^2 \vev{i-1~\ell_3}[\ell_3~\ell_2]\vev{\ell_2~i-1} &=
K_{41}^2 K_{12}^2- K_1^2 K_3^2 \cr &=(p_{i-1}+K_1)^2 (K_1+K_2)^2
}}
we can read the coefficient from part (a) as
\eqn\coeffa{ {(K_2^2)^3\vev{i-3~i-2}\vev{i-2~i-1}^4
\vev{i-1~j}^4\vev{i+r-1~i+r}/\prod_{s=1}^{n} \vev{s~s+1}\over
\vev{ i+r-1| K_2\cdot p_{i-2}|i-1 }\vev{ i+r| K_2\cdot  p_{i-2}|i-1 }
\vev{i-3| K_2\cdot K_1| i-1}\vev{i-2| K_2\cdot K_1| i-1}}
}

For part (b) we must use different spinor relations,
$\lambda_{\ell_4}=\tilde{\a} \lambda_{i-2}$,
$\lambda_{\ell_3}=\tilde {\b} \lambda_{i-2}$ and
an analog of the identity \bigbux, namely
\eqn\idenb{\eqalign{
{ [\ell_4~\ell_3]^3 \over [\ell_3~i-2][i-2~\ell_4]}
\tilde{\a}^2\tilde {\b}^2
\vev{i-2~\ell_2}[\ell_2~\ell_1]\vev{\ell_1~i-2} & =
K_{34}^2 K_{41}^2- K_4^2 K_2^2 \cr & =(p_{i-2}+p_{i-1})^2 (p_{i-1}+K_1)^2,
}}
to read out the coefficient
\eqn\coeffb{{\vev{j|K_1\cdot K_2|i-2}^4 \vev{i-2~i-1}^4
\vev{i-1~i}\vev{i+r-1~i+r}/(K_1^2 \prod_{s=1}^{n} \vev{s~s+1})
\over \vev{i-1|K_1\cdot K_2|i-2} \vev{i|K_1\cdot K_2|i-2}
\vev{i+r-1|K_1\cdot  p_{i-1}|i-2}\vev{i+r|K_1\cdot  p_{i-1}|i-2}}.
}

Putting it all together, we finally have
\eqn\totalb{\eqalign{& d_{n:r;i} = \cr &
{(K_2^2)^3\vev{i-3~i-2}\vev{i-2~i-1}^4
\vev{i-1~j}^4\vev{i+r-1~i+r}/\prod_{s=1}^{n} \vev{s~s+1}\over
\vev{ i+r-1| K_2\cdot p_{i-2}|i-1 }\vev{ i+r| K_2\cdot  p_{i-2}|i-1 }
\vev{i-3| K_2\cdot K_1| i-1}\vev{i-2| K_2\cdot K_1| i-1}}\cr
& + {\vev{j|K_1\cdot K_2|i-2}^4 \vev{i-2~i-1}^4
\vev{i-1~i}\vev{i+r-1~i+r}/(K_1^2 \prod_{s=1}^{n} \vev{s~s+1})
\over \vev{i-1|K_1\cdot K_2|i-2} \vev{i|K_1\cdot K_2|i-2}
\vev{i+r-1|K_1\cdot  p_{i-1}|i-2}\vev{i+r|K_1\cdot  p_{i-1}|i-2}}
}}
We have compared this general formula with coefficient $c_{457}$
of seven gluons in the  helicity configuration $(1^-2^-3^+4^-5^+6^+7^+)$ in \BernKY\
and found
that they agree.

\newsec{Summary}

In this paper we have 
reduced the problem of computing the coefficient of any scalar box integral in any one-loop $\N=4$ amplitude to finding solutions to the four equations
\eqn\nullsheis{
\ell^2=0, \quad (\ell-K_1)^2=0, \quad (\ell
-K_1-K_2)^2=0, \quad (\ell+K_4)^2=0}
in $(--++)$ signature.
From this set of solutions it is possible to read off the coefficient from the formula
\eqn\besatwaydi{{\hat a}_\alpha = {1 \over |{\cal S}|}\sum_{{\cal S},J}  n_J A^{\rm tree}_1 A^{\rm tree}_2 A^{\rm tree}_3 A^{\rm tree}_4.
}
It would be interesting to perform a full classification of helicity configurations to obtain explicit formulas for all coefficients to all multiplicities.

\bigskip
\bigskip
\centerline{\bf Acknowledgments}

We thank L. Dixon and E. Witten for helpful conversations.
R. B. and B. F. were supported by NSF grant PHY-0070928. F. C. was
supported in part by the Martin A. and Helen Chooljian Membership
at the Institute for Advanced Study and by DOE grant
DE-FG02-90ER40542.

\appendix{A}{Discontinuities of the Four-Mass Scalar Box Integral}

In this section we study in detail the discontinuities of the
four-mass box integral. The motivation is to allow us to perform
non-trivial consistency checks on the coefficients found in the
previous section and to discuss the difficulties in trying to
compute them by the application of coplanar operators along the
lines of \refs{\CachazoDR,\BrittoNJ}.

One way to obtain the discontinuities is to compute the imaginary
part of the explicit formulas for the four-mass integral in a
kinematical regime chosen in order to isolate the cut of interest.
Another way, which we find more intuitive, is to compute the
discontinuity of the scalar box integral by cutting
it directly. It turns out that the cut integral is easy to evaluate
explicitly, as we now describe.

\subsec{Cut Integral}

Consider first the four-mass box function,
\eqn\ifm{ I^{4m}= \int d^4\ell { 1\over (\ell^2+i\epsilon) (
(\ell-K_1)^2+i\epsilon) ( (\ell -K_1-K_2)^2+i\epsilon)
((\ell+K_4)^2+i\epsilon) }.}

Let us start with the discontinuity in the $(K_1, K_2)$-channel.
The integral we have to evaluate is obtained from \ifm\ by
replacing the first and third propagators by the delta functions
imposing the on-shell condition.

\eqn\icut{\Delta I^{4m} |_{K_{12}^2>0} = \int d^4\ell { \delta^{(+)}(\ell^2)
\delta^{(+)}((\ell-K_1-K_2)^2) \over (\ell-K_1)^2 (\ell+K_4)^2}.}

Now we can parameterize $\ell_{a\dot a} =
t\lambda_a\tilde\lambda_{\dot a}$ and write the Lorentz invariant
measure as follows:
\eqn\lim{\Delta I^{4m} |_{K_{12}^2>0} =  \int_0^{\infty}t~ dt \int
\vev{\lambda~d\lambda}[\tilde\lambda ~ d\tilde\lambda ]
{\delta^{(+)}((\ell-K_1-K_2)^2) \over (\ell-K_1)^2 (\ell+K_4)^2}.}

The remaining delta function can be written as
\eqn\arg{\delta((\ell-K_1-K_2)^2) = \delta( t
\lambda_a\tilde\lambda_{\dot a}K_{12}^{a\dot a} - K_{12}^2) =
{1\over \lambda_a\tilde\lambda_{\dot a}K_{12}^{a\dot a}} \delta
(t- {K_{12}^2\over \lambda_a\tilde\lambda_{\dot a}K_{12}^{a\dot
a}}
 ),}
where $K_{12}=K_1+K_2.$
Therefore the $t$ integral can be performed to find
\eqn\tin{ \Delta I^{4m} |_{K_{12}^2>0} = \int \vev{\lambda~d\lambda}[\tilde\lambda ~
d\tilde\lambda ] {K_{12}^2 \over (\lambda_a\tilde\lambda_{\dot
a}K_{12}^{a\dot a})^2(\ell-K_1)^2 (\ell+K_4)^2}. }
The denominator can be written as
\eqn\den{(\lambda_a\tilde\lambda_{\dot a}K_{12}^{a\dot
a})^2(\ell-K_1)^2 (\ell+K_4)^2 = (Q^{a\dot
a}\lambda_a\tilde\lambda_{\dot a}) (S^{a\dot
a}\lambda_a\tilde\lambda_{\dot a}),}
with
\eqn\defsq{ \eqalign{ Q^{a\dot a} = & ~ -K_{12}^2 K_1^{a\dot a} +
K_1^2 K_{12}^{a\dot a}, \cr S^{a \dot a} = & ~K_{12}^2 K_4^{a\dot
a} + K_4^2 K_{12}^{a\dot a}. } }

The way to evaluate this integral is to use a Feynman
parametrization.  This gives
\eqn\fey{\Delta I^{4m} |_{K_{12}^2>0} =  \int_0^1 d\alpha_1d\alpha_2 \delta (\alpha_1+\alpha_2-1)
 \int \vev{\lambda~d\lambda}[\tilde\lambda ~
d\tilde\lambda ]{1\over (P^{a\dot a}\lambda_a \tilde \lambda_{\dot
a})^2},}
with
\eqn\deP{ P^{a\dot a} = \alpha_1 Q^{a\dot a} + \alpha_2 S^{a\dot
a}.}

The integral over the sphere was evaluated in \CachazoKJ\ and gives
$1/P^2$. Therefore we have
\eqn\fini{\Delta I^{4m} |_{K_{12}^2>0} =   \int_0^1 d\alpha_1d\alpha_2 \delta
(\alpha_1+\alpha_2-1){1\over \alpha_1^2 Q^2 + 2 Q\cdot S\;\alpha_1
\alpha_2+ \alpha_2^2 S^2}. }
Now we can perform the integration in $\alpha_1$ and $\alpha_2$.
The result is
%
\eqn\wivx{\Delta I^{4m} |_{K_{12}^2>0} =  {1 \over \rho K_{41}^2 K_{12}^2}
\log{\left({ (1-\lambda_1-\lambda_2-\rho)^2 \over 4 \lambda_1 \lambda_2}\right)},}
where $\rho$ is defined as in \defrho.
We can similarly evaluate the discontinuity of the four-mass box function in the $K_1$-channel.  This is obtained from \ifm\ by replacing the first and second propagators by the delta functions imposing the on-shell condition:
\eqn\ccut{\Delta I^{4m} |_{K_{1}^2>0} =  \int d^4\ell { \delta^{(+)}(\ell^2)
\delta^{(+)}((\ell-K_1)^2) \over (\ell-K_1-K_2)^2 (\ell+K_4)^2}.}
Again, parametrize $\ell_{a\dot a} =
t\lambda_a\tilde\lambda_{\dot a}$ and then perform the $t$ integral to find
\eqn\ctin{\Delta I^{4m} |_{K_{1}^2>0} =  \int \vev{\lambda~d\lambda}[\tilde\lambda ~
d\tilde\lambda ] {K_{1}^2 \over (\lambda_a\tilde\lambda_{\dot
a}K_{1}^{a\dot a})^2(\ell-K_1-K_2)^2 (\ell+K_4)^2}. }
Here the denominator can be written as
\eqn\cden{(\lambda_a\tilde\lambda_{\dot a}K_{1}^{a\dot
a})^2(\ell-K_1-K_2)^2 (\ell+K_4)^2 = (Q^{a\dot
a}\lambda_a\tilde\lambda_{\dot a}) (S^{a\dot
a}\lambda_a\tilde\lambda_{\dot a}),}
with
\eqn\cdefsq{ \eqalign{ Q^{a\dot a} = & ~ K_{12}^2 K_1^{a\dot a} -
K_1^2 K_{12}^{a\dot a}, \cr S^{a \dot a} = & ~K_{1}^2 K_4^{a\dot
a} + K_4^2 K_{1}^{a\dot a}. } }
So we may follow exactly the same steps as in the previous case, and here we find the result
\eqn\cwivx{\Delta I^{4m} |_{K_{1}^2>0} =  -{1 \over \rho K_{41}^2 K_{12}^2}
\log{\left({ (1-\lambda_1+\lambda_2-\rho)^2 \over 4 \lambda_2}\right)},}
where $\lambda_1,\lambda_2,\rho$ are the same expressions defined in \defrho.

\subsec{Imaginary Part}

As a consistency check, we can reproduce our results \wivx\ and \cwivx\ from
the imaginary part of the explicit form of the four-mass integral
in the appropriate kinematical regimes.

The integral \ifm\ was computed explicitly in \DennerQQ\ and is given in
terms of the invariants $\K_{ml} = -(K_m + K_{m+1} + \ldots +
K_{l-1})^2$ as follows:
\eqn\expi{\eqalign{I^{4m} &= {1\over a (x_1-x_2)}\sum_{j=1}^2 (-1)^j
\left( -\half \ln^2(-x_j) \right. \cr &   -{\rm Li}_2\left( 1+
{\K_{34}-i\epsilon \over \K_{13}-i\epsilon }x_j \right)
-\eta\left( -x_k, {\K_{34}-i\epsilon \over \K_{13}-i\epsilon }
\right) \ln \left( 1+ {\K_{34}-i\epsilon \over \K_{13}-i\epsilon
}x_j \right) \cr & -{\rm Li}_2\left( 1+ {\K_{24}-i\epsilon \over
\K_{12}-i\epsilon }x_j \right) -\eta\left( -x_k,
{\K_{24}-i\epsilon \over \K_{12}-i\epsilon } \right) \ln \left( 1+
{\K_{24}-i\epsilon \over \K_{12}-i\epsilon }x_j \right) \cr &
\left. +\ln(-x_j)(\ln(\K_{12}-i\epsilon ) + \ln(\K_{13}-i\epsilon
) - \ln(\K_{14}-i\epsilon ) - \ln(\K_{23}-i\epsilon )  ) \right). }
}
Here the function $\eta(x,y)$ is given by
\eqn\etabranch{\eta(x,y)=2\pi i [\vartheta(-\Im x)\vartheta(-\Im
y)\vartheta(\Im(xy)) -\vartheta(\Im x)\vartheta(\Im
y)\vartheta(-\Im(xy))], }
and $x_1$ and $x_2$ are the roots of a quadratic polynomial:
\eqn\quoio{a x^2+b x+ c + i\epsilon d = a (x-x_1)(x-x_2),}
with
\eqn\defss{\eqalign{ &  a = \K_{24}\K_{34}, \cr & b=
\K_{13}\K_{24}+ \K_{12}\K_{34}-\K_{14}\K_{23}, \cr & c =
\K_{12}\K_{13}, \cr & d = \K_{23}. }}

The $i\epsilon$ prescription in \expi\ allows us to use this
formula in any kinematical regime. The main simplification is that
the proper branch of each of the functions in \expi\ is simply
given by the principal branch. The formula for $I^{4m}$ presented
in \BernKR\ can be recovered from \expi\ by setting the $\eta$
functions and $\epsilon$ to zero. Even though the formula looks
simpler in that form, it has the disadvantage that it is not clear
which branch of the various dilogarithms and logarithms compute
the appropriate discontinuity.

\appendix{B}{Twistor Space Structure of Four-Mass Box Coefficients}

In this section we study the twistor space localization of the
four-mass scalar box function coefficient found in section 3.1.
The way we choose to do this can also be thought of as a
consistency check on the coefficient.

Consider for example the amplitude
$A_{8;1}(1^-,2^-,3^-,4^-,5^+,6^+,7^+,8^+)$. The two four-mass box
integrals have momenta distributed as $(12)-(34)-(56)-(78)$ and
$(23)-(45)-(67)-(81)$. In the first case we find that the
coefficient must be zero, since one of the tree-level amplitudes
in the quadruple cut diagram is necessarily zero. This is
essentially equivalent to the observation of \BernKY, where the
same conclusion was derived from a triple cut.

In order to study the second four-mass integral, consider the cut
in the $(2345)$-channel. This cut is given by
\eqn\cutff{ C_{2345} = \int d\mu~ A^{\rm tree}_6(\ell_1 , 2^-, 3^-,
4^-, 5^+, \ell_2)A^{\rm tree}_6(-\ell_2 , 6^+, 7^+, 8^+
,1^-,-\ell_1) .}

There are four contributions to this cut depending on the
helicities $(h_{\ell_1},h_{\ell_2})$ of the particles
  running in the cut propagators. If
$(h_{\ell_1},h_{\ell_2}) =(-,-)$ then we get zero. If
$(h_{\ell_1},h_{\ell_2}) =(-,+)$ or $(h_{\ell_1},h_{\ell_2})
=(+,-)$, then the whole $\N=4$ supermultiplet contributes. In this
case, both tree-level amplitudes are very simple; they become a
mostly-minus MHV and a mostly-plus MHV, respectively. These
contributions can be easily written as
\eqn\sift{ {\gb{ 1|(2+3+4)|5}^4 \over ((p_1+p_2+p_3+p_4)^2)^4}
\left[ C_{1234}\right]_{i\rightarrow i+1}.}

Since the four-mass integral does not contribute to the cut
$C_{1234}$, then it does not contribute to this part of the cut
$C_{2345}$ either.

Finally, we have the case with $(h_{\ell_1},h_{\ell_2}) =(+,+)$.
In this case, only gluons can propagate; the complication
arises from the fact that both tree-level amplitudes are next-to
MHV six-gluon amplitudes.

It turns out that we cannot use collinear operators to extract
information from this cut. The reason is that a collinear operator
does not localize the integral. We need a coplanar operator. 

A coplanar operator is defined as follows \WittenNN.
\eqn\hereisk{K_{ijkl}= \vev{i~j}[\tilp_k~\tilp_l]+
\vev{j~k}[\tilp_i~\tilp_l] + \vev{k~i}[\tilp_j~\tilp_l]+
\vev{k~l}[\tilp_i~\tilp_j] + \vev{i~l}[\tilp_j~\tilp_k]+
\vev{j~l}[\tilp_k~\tilp_i],}
where
\eqn\ouros{ (\tilp_i)_\da = {\del \over {\del \lt^\da_i}}.}

In
this case we could use $K_{2345}$ or $K_{6781}$ to produce
rational functions.

Recall that the cut can also be written as the discontinuity of
the amplitude in the $(2345)-$channel.
\eqn\disc{ C_{2345} = \Delta A_8^{\rm 1-loop} = \ldots + {\hat
f}\Delta I^{4m}.}
We have only explicitly written the term from the four-mass
integral. Once we apply either of the two coplanar operators $K$ on
the cut integral, we produce a rational function. On the other
hand, acting with $K$ on \disc\ produces
\eqn\prod{ K C_{2345} = \ldots + K ( {\hat f} \Delta I^{4m}).}

Following the arguments of \CachazoDR, one can show that the terms in the
ellipses, which come from $1m$, $2m$ and $3m$ scalar box
integrals, are rational functions. Therefore we conclude that
\eqn\con{ K ( {\hat f} \Delta I^{4m} ) }
must be rational.

But we can go even farther. In the previous section, we found that
\eqn\prew{\Delta I^{4m} = {1\over \rho K_{41}^2 K_{12}^2 }\log
\left( {(1-\lambda_1 - \lambda_2 -\rho)^2 \over 4
\lambda_1\lambda_2} \right). }
Therefore, when none of the derivatives in $K$ act on the
logarithm, we find a term of the form
\eqn\loqw{ K\left( {{\hat f}\over \rho K_{41}^2 K_{12}^2} \right)
\times \log \left( {(1-\lambda_1 - \lambda_2 -\rho)^2 \over 4
\lambda_1\lambda_2} \right).}

The only way this is consistent with \con\ being rational is that
\eqn\twist{ K\left( {{\hat f}\over \rho K_{41}^2 K_{12}^2} \right)
= 0.}

Recall that $f = -2 {\hat f}/ (\rho  K_{41}^2 K_{12}^2)$ is the
definition of the four-mass box function coefficient, which we
claimed would have a simple twistor structure configuration.
Indeed, \twist\ is the reason for our claim.

So we have found that $K_{2345} f_2 = K_{6781} f_2 =0$. Note that this
four-mass box integral also has a cut in the $(4567)-$channel.
Using the same logic we find that $K_{4567} f_2 = K_{8123} f_2
=0$.

The conclusion is then that all gluons in two adjacent corners of
the four-mass box function coefficient are localized on a plane.
The most general configuration consistent with this picture is
shown in Figure 13.

\ifig\fourlines{The twistor space structure of the four-mass coefficient $f_2$.
The four lines are not all coplanar.
} {\epsfxsize=0.50\hsize\epsfbox{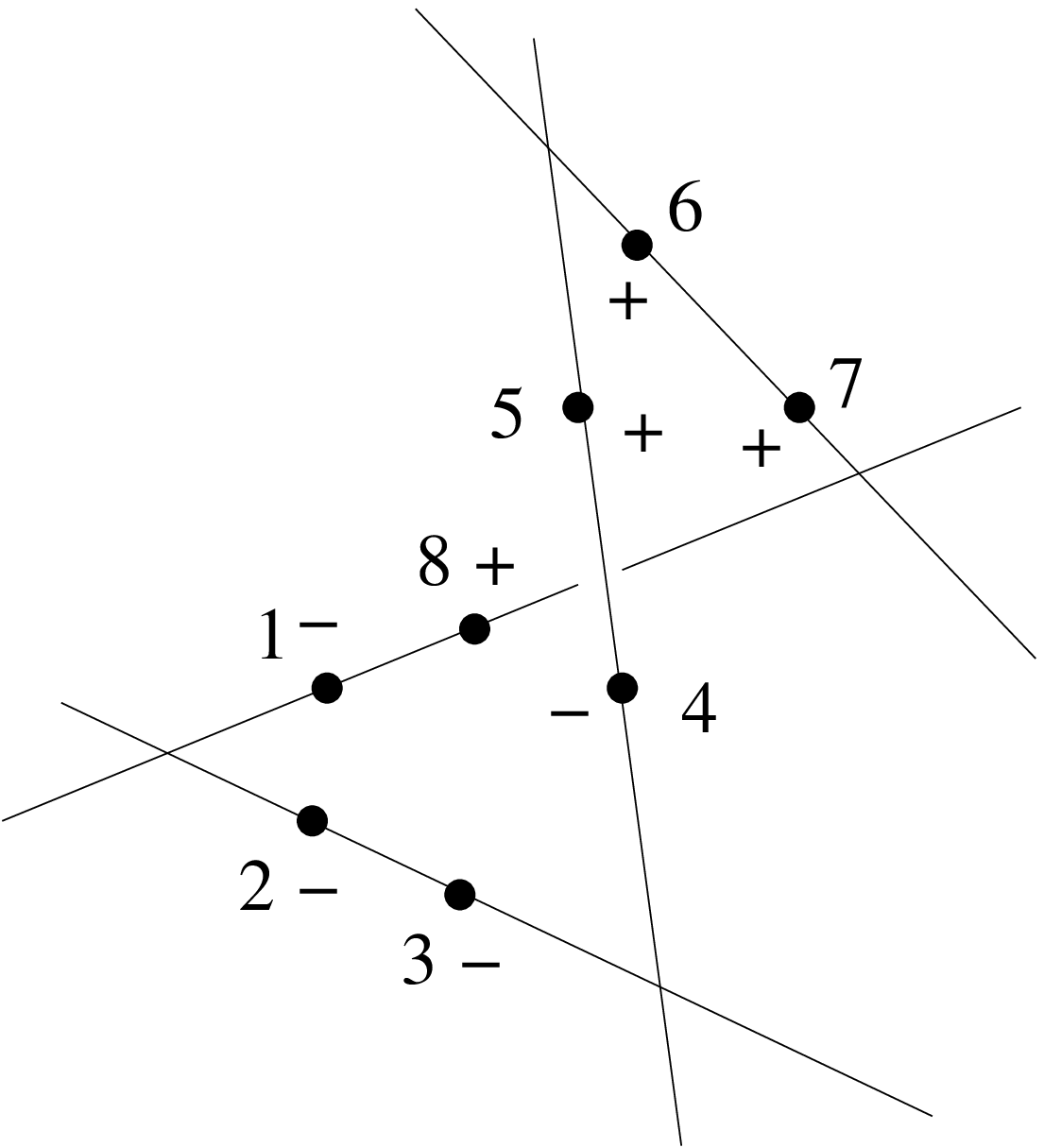}}

The consistency checks we have run on the coefficient found in
section 3.1 are the following: $f_2$ is annihilated by the
coplanar operators mentioned before, and \con\ is a rational
function.

At this point we must comment on why studying this cut does not
give an efficient way of computing $f_2$. After all, computing the
rational function $K_{2345} C_{2345}$ from the cut integral
representation of $C_{2345}$ is very simple. Moreover, it turns out that
all other
coefficients contributing to this cut can be computed from
cuts in three-particle channels, along with linear equations from the infrared singularities, using the method of \BrittoNJ.
 Therefore, all reduces to
\eqn\reda{ W = K ( {\hat f} \Delta I^{4m} ), }
where $W$ is a known rational function, and $\Delta I^{4m}$ is also
known.

Let us write $K$ schematically as ${\cal O}_1{\cal O}_2$. Then,
expanding the right hand side of \reda\ we find
\eqn\readg{ W = K(\hat f) \Delta I^{4m} + {\cal O}_1(\hat f){\cal
O}_2(\Delta I^{4m})  + {\cal O}_2(\hat f){\cal O}_1(\Delta I^{4m})
+ {\hat f} K (\Delta I^{4m} ).}

The problem arises when one notices that the only term that has no
derivatives acting on ${\hat f}$ vanishes. Indeed, it not
difficult to show with the help of a symbolic manipulation program
that $K (\Delta I^{4m}) =0$.

\listrefs

\end